\documentclass[aps,12pt,nofootinbib,showpacs,showkeys,preprintnumbers,amsmath,amssymb]{revtex4-1}		

\usepackage{amsmath,amsfonts,amssymb,amsthm,bm,bbm,bbold,braket,color,dsfont,fancybox,hyperref,
srcltx,subcaption,graphicx,ucs,yfonts,cancel,xfrac,verbatim,xcolor}
        
\usepackage{float}

\usepackage[section]{placeins}

\usepackage{lineno}

\begin{document}

\title{Probing Left-handed Heavy Neutral Leptons in the Vector Scotogenic Model}

\author{Paulo Areyuna C.$^{1,3}$}
\email{paulo.areyuna@sansano.usm.cl}

\author{Jilberto Zamora-Saa$^{2,3}$}
\email{jilberto.zamora@unab.cl} \email{jilberto.zamorasaa@cern.ch}

\author{Alfonso R. Zerwekh$^{1,3,4}$}
\email{alfonso.zerwekh@usm.cl}

\affiliation{$^1$ Universidad Técnica Federico Santa María Casilla 110-V, Valparaíso, Chile.}
\affiliation{$^2$Center for Theoretical and Experimental Particle Physics - CTEPP, Facultad de Ciencias Exactas, Universidad Andres Bello, Fernandez Concha 700, Santiago, Chile.}
\affiliation{$^3$Millennium Institute for Subatomic physics at high
energy frontier - SAPHIR, Fernandez Concha 700, Santiago, Chile.}
\affiliation{$^4$Centro Científico - Tecnológico de Valparaíso, Casilla 110-V, Valparaíso, Chile}

\begin{abstract}
In this work, we consider an extension to the Standard Model composed by a Massive Vector Doublet under SU(2)$_L$ and a Left-handed Heavy Neutral Lepton. We study the production of these exotic leptons with the Same Flavor Opposite Sign standard lepton pair, and jets, considering Drell-Yan and Vector Boson Fusion as independent cases. We find that for the latter, the dilepton angular distribution is different enough from the background to use it as a smoking-gun for our model. Based on this fact, we establish limits on the parameter space considering previous experimental searches in this final state. 
\end{abstract}

\maketitle

\section{Introduction}\label{sec:intro}
The standard model of particle physics (SM) is the most complete and successful explanation of the fundamental interactions. Despite its awesome phenomenological achievements, this model is not able to explain all the observed phenomena. Among them we can mention, for example, the presence of Dark Matter (DM) in the universe. Although  we know very well the amount of DM  in the Universe (in terms of relic density) \cite{PLANCK}, we know nothing about it's nature.\newline
Another problem is the neutrino mass generation mechanism. Due to the apparent absence of right-handed neutrinos, it is impossible to generate neutrino masses through Yukawa couplings as the other fermions do. Moreover, even if right handed neutrinos exist, the Yukawa coupling needed to explain the neutrino masses would be so small that the standard mechanism for mass generation seems to be unnatural for the neutrinos. In this context, it would be of great interest if these two apparently independent problems (dark matter and neutrino mass generation) could be connected. Part of the aim of this work is to propose a mechanism for studying one realization of this intriguing possibility.

Among the variety of proposed ideas that may describe the physics of dark matter, the minimal  dark matter program  \cite{minimaldm} appears as specially appealing. Initially, it was developed for scalar and fermion fields. However, it was later shown that the mechanism worked very well also for vector fields. In reference \cite{triplet_dm}, for instance, it was shown that the inclusion of a new massive field in the adjoint representation of $SU(2)_L$ provide a viable dark matter candidate. Additionally, the case of a massive vector field in the fundamental representation of $SU(2)_L$ was also successfully studied in reference \cite{vector_dm}. Interestingly enough, this last model can be naturally extended by the inclusion of a sterile left-handed heavy neutral lepton (HNL) \cite{masses_and_mixings}. In this version, the model is capable of solving both problems: dark matter and neutrino mass generation mechanism \cite{vector_dm,masses_and_mixings,dong2021}. Indeed, this model has been shown to be able of tackling also the muon $(g-2)$ problem \cite{dong2021}.

In this work, we perform a phenomenological study for the production of these new particles ({\it i.e.} the new massive vector boson and the left-handed HNL)  in the Large Hadron Collider (LHC). We use existing experimental data to constrain the model parameters, as a function of the sensitivity of current experiments. Additionally, we make predictions based on the kinematical features of the production process, in order to facilitate the search of these new particles in future measurements.

The paper is organized as follows: In Section \ref{sec:theo} we describe the main features of the proposed model, while in Section \ref{sec:prod}  we discuss the main production mechanisms of these new particles at the LHC. In Sections \ref{sec:dy} and \ref{sec:vbf} we study the production mechanisms in order to define upper limits on the parameter space, and focus on the kinematical signatures of each process. In Section \ref{sec:proj} we study the discovery potential of the model at future experimental facilities, and in Section \ref{sec:lowlims} we obtain an estimation of lower limits for the parameter space coming from dark matter measurements. Finally, our conclusions are presented in Section \ref{sec:conc}.

\section{Theoretical background}\label{sec:theo}
Let us start by introducing the following set of vectors:
\begin{equation}
    V_{\mu}=\begin{pmatrix}V^{+}_\mu\\V^0_\mu\end{pmatrix}=\begin{pmatrix}V^{+}_\mu\\\frac{1}{\sqrt{2}}(V^1_\mu+iV^{2}_\mu)\end{pmatrix},
\end{equation}
which transforms as $(1,2,1/2)$ under the SM group $SU(3)_c\times SU(2)_L\times U(1)_Y$. The dynamics of this new vector doublet is described by the following lagrangian:
\begin{equation}
\begin{split}
    \mathcal{L}=\mathcal{L}_{SM}&-\frac{1}{2}(D_\mu V_\nu- D_\nu V_\mu)^\dagger (D^\mu V^\nu -D^\nu V^\mu)
    +i\frac{g'\kappa_1}{2}V_\mu^\dagger B^{\mu\nu}V_\nu+ig\kappa_2 V_\mu^\dagger W^{\mu\nu}V_\nu \\
    &+M_V^2 V_\mu^\dagger V^\mu-\alpha_2(V_\mu^\dagger V^\mu)(V_\nu^\dagger V^\nu)-\alpha_3(V_\mu^\dagger V^\nu)(V_\nu^\dagger V^\mu)\\&-\lambda_2(\Phi^\dagger \Phi)(V_\mu^\dagger V^\mu)-\lambda_3(\Phi^\dagger V_\mu)(V^{\mu\dagger}\Phi)-\frac{\lambda_4}{2}[(\Phi^\dagger V_\mu)(\Phi^\dagger V^\mu)+(V^{\mu\dagger}\Phi)(V_\mu^\dagger \Phi)],
    \end{split}
\end{equation}
where $D_\mu$ stands for the covariant derivative, $W^{\mu\nu}$ and $B^{\mu\nu}$ are the field strengths of $SU(2)_L$ and $U(1)_Y$, respectively, and $\Phi$ is the SM Higgs doublet. This lagrangian was labeled by the authors as the Vector Doublet Dark Matter Model (VDDMM), since 
the neutral component of the vector has been proven to account for DM (see reference \cite{vector_dm}). 
It's worth to mention the presence of non minimal gauge interactions. The coupling constants associated to these interactions ($\kappa_1$ and $\kappa_2$) are free parameters. In reference \cite{vector_dm} the authors set their values as $\kappa_1=\kappa_2=1$, in order to avoid interactions between the photons and the neutral component of the vector doublet.
This restriction can be full filled with a more general condition: $\kappa_1=\kappa_2=\kappa$. However, since the phenomenology of the model has been studied with $\kappa=1$, we are going to keep this choice, despite the fact that a different choice of this value can produce an interesting phenomenology. Additionally, the lagrangian presents an accidental $Z_{2}$ symmetry, making the lightest neutral vector as a viable DM candidate.
\newline

\begin{figure}[!h]
    \centering
    \includegraphics[width=0.5\textwidth]{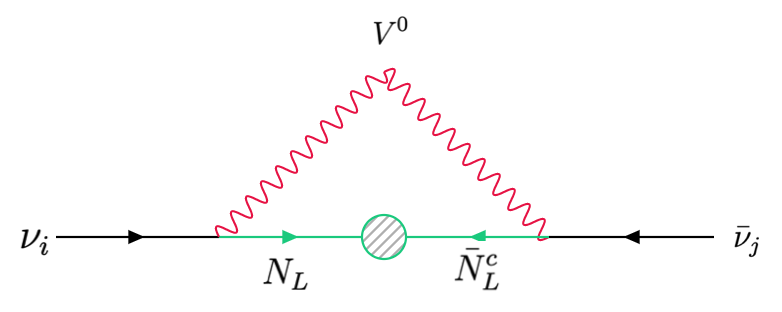}
    \caption{Neutrino mass generation mechanism in the Vector Scotogenic Model.}
    \label{neutrinomass}
\end{figure}

In order to perform calculations, it's useful to define some quantities after the symmetry breaking:
\begin{equation}
    \begin{split}
        &M_{V^\pm}^2=\frac{1}{2}(2M_V^2-v^2\lambda_2)\\
        &M_{V^1}^2=\frac{1}{2}(2M_V^2-v^2[\lambda_2+\lambda_3+\lambda_4])\\
        &M_{V^2}^2=\frac{1}{2}(2M_V^2-v^2[\lambda_2+\lambda_3-\lambda_4])\\
        &\lambda_L=\lambda_2+\lambda_3+\lambda_4.
    \end{split}
\end{equation}
The latter is the coupling that describes the interaction between the Higgs boson and $V^1$. Moreover, in this work, we are particularly interested on the production of these vector states, therefore, we don't need to pay attention on the quartic interactions, having in consequence  only 4 relevant parameters. For convenience, we perform a change of basis for them, and write the lagrangian in terms of the previously presented parameters:
\begin{equation}
\begin{split}
    &\lambda_2=\lambda_L+2\frac{M_{V^1}^2-M_{V^\pm}^2}{v^2}\\
    &\lambda_3=\frac{2M_{V^\pm}^2-M_{V^1}^2-M_{V^2}^2}{v^2}\\
    &\lambda_4=\frac{M_{V^2}^2-M_{V^1}^2}{v^2}\\
    &M_V^2=M_{V^1}^2+\frac{v^2}{2}\lambda_L.
    \end{split}
\end{equation}
Henceforth, we are going to consider the special case of $M_{V^1}= M_{V^2}\equiv M_{V^0}$. This special case is motivated by the results obtained in reference \cite{vector_dm}. In addition, we define the quantity $\Delta M= M_{V^+}-M_{V^0}$.\newline
Due to the vector doublet quantum numbers and Lorentz invariance, it is not possible to link directly the new vector field to the SM fermions. However, if we introduce an exotic left-handed neutrino which is singlet of the SM, we can write the following interaction term:
\begin{equation}
    \mathcal{L}_{VNL}=-\sum_{k=\{e,\mu,\tau\}}\beta_k \bar{L}_k \gamma^\mu \tilde{V_\mu} N_L  +\text{h.c.} \ ,
    \label{LVNL}
\end{equation}
with
\begin{equation}
    \tilde{V_\mu}=i\sigma_2 V_\mu^*=\begin{pmatrix}\frac{1}{\sqrt{2}}(V^1_\mu-iV^{2}_\mu)\\-V^{-}_\mu\end{pmatrix} \ .
\end{equation}
In Eq.~\ref{LVNL}, $N_L$ stands for a left-handed Majorana fermion with definite mass $M_N$, which acts as a portal between the vector doublet and SM leptons. If we define $N_L$ as $Z_2$ odd, the accidental symmetry is respected, and this new fermion can also be a DM candidate, depending if it's lighter than the vector candidate. The addition of a new fermion state allows the generation of radiative neutrino masses (see Figure \ref{neutrinomass}). The necessary conditions for reproducing the neutrino mixing matrix have been studied with two \cite{masses_and_mixings} and three \cite{dong2021} HNLs. Since the scope of the present work is focused on determining the detection prospects of the model at the LHC, we will work with just one HNL. This simplifying assumption allows us to explore different values of the $\beta$ couplings without concerning about neutrino mass generation,  because it is always possible to attribute the neutrino mass generation to extra HNLs with masses beyond LHC sensitivity. Due to the similarities between the scotogenic model proposed by Ma in reference \cite{Ma:2006km} and our construction, we will refer to the model treated in this work as the Vector Scotogenic Model. 

\section{Production mechanism at the LHC}\label{sec:prod}

\begin{figure}[!h]
    \begin{subfigure}{0.45\textwidth}
    \includegraphics[width=\textwidth]{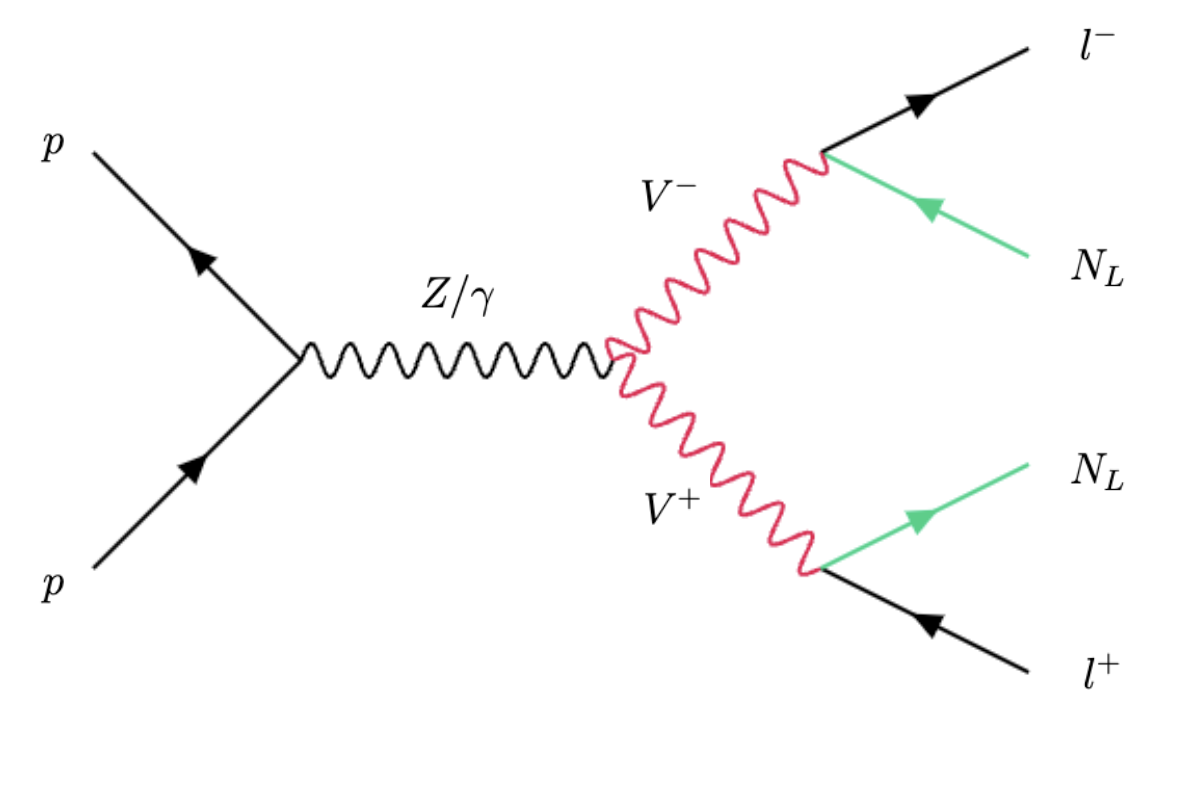}
    \caption{Drell-Yan contribution}
\end{subfigure}
    \begin{subfigure}{0.45\textwidth}
    \includegraphics[width=\textwidth]{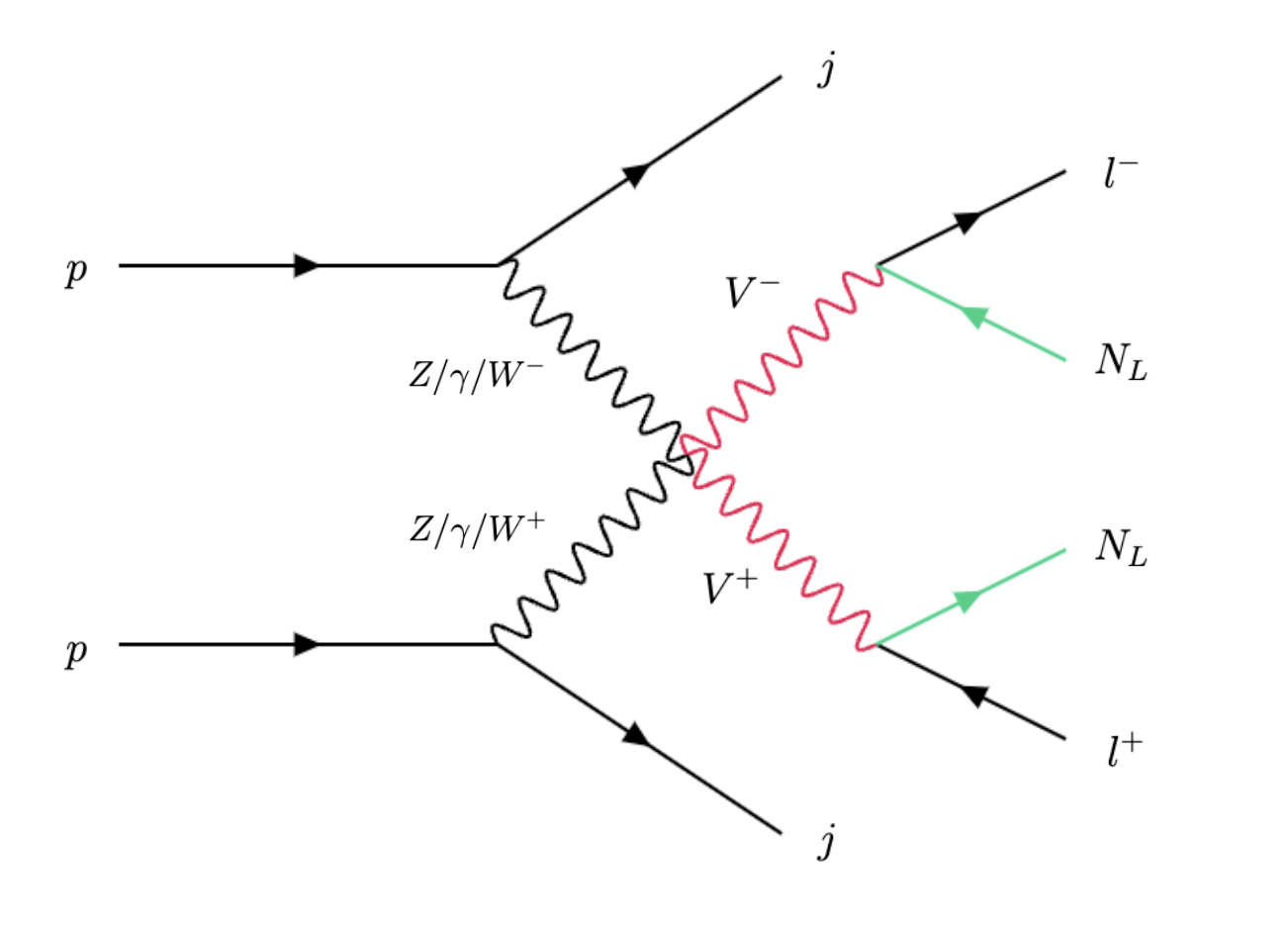}
    \caption{Vector Boson Fusion contribution}
\end{subfigure}

\caption{Production mechanisms at the LHC.}
\label{mechanisms}
\end{figure}

As we stated above, we are interested in the study of the Vector Scotogenic model in the context of LHC, particularly at the ATLAS detector. The signal to be searched is composed by two HNL in the final state, which can be used as smoking gun of this new physics. These fermions can be produced mainly in two type of processes: Drell-Yan (DY) and Vector Boson Fusion (VBF) as can be seen in Figure \ref{mechanisms}. The main difference between these mechanisms is the presence of jets in the final state of VBF.

 We used FeynRules  \cite{fr1,fr2,ufo} to obtain the Feynman rules for the new sector, and Madgraph5\_aMC@NLO version 3.5.0  \cite{mg5} to compute the cross section for these processes at $\sqrt{s}=13$ [TeV], considering all the possible diagrams.
It's worth mentioning that the production process depends strongly on the charged vector decay width. Since we are restricted to the completely degenerate scenario, the only decay channel is through the trilinear term in eq. \eqref{LVNL}. Then, neglecting the lepton masses, this quantity can be written as follows
\begin{equation}
    \Gamma_{V^+}=(|\beta_e|^2 + |\beta_\mu|^2+|\beta_\tau|^2)\frac{ (M_{V^+}^2 - M_N^2)^2   (2 M_{V^+}^2 +M_N^2) }{(48 \pi M_{V^+}^5)}.
\end{equation}
We have performed our analysis for different values of $M_{V^+}$ and $\beta_\mu$, restricting ourselves to the degenerate case where $\Delta M=0$. As a first step, we have considered only muons in the final state. The study was carried out considering fixed values for the remaining parameters, presented in Table \ref{simulation_params}. The choice of the $\beta$ couplings has been motivated by the strong constraints on Lepton Flavor Violating decays. 
\begin{table}[!h]
    \centering
    \begin{tabular}{|c|c|c|c|c|}
    \hline
            & $M_N$ &  $\beta_e$&$\beta_\tau$&  $\lambda_L$ \\
       \hline
        BP1 &$50 [$GeV$]$& $0$ & $0.5$ & $ 5$\\
        BP2 &$50 [$GeV$]$& $0$ & $1$ & $ 5$\\

        \hline
    \end{tabular}
    \caption{Benchmark values for the fixed parameters.}
    \label{simulation_params}
\end{table}
\newline
 Additionally, we have simulated the SM background for each type of the studied processes at $\sqrt{s}=13$ [TeV]: $p p  \rightarrow  \mu^+  \mu^- \bar{\nu}\nu$ for 
 DY and $p p  \rightarrow  j j \mu^+  \mu^- \bar{\nu}\nu$ for VBF
 (notice that in both cases the neutrinos can be of any flavour).

 Firstly, we have computed the cross section for each process without considering kinematical cuts and the detector efficiency, as can be seen in Table \ref{bench_prel}. Here, we have found that DY cross sections dominate by one order of magnitude over VBF. In addition, the current experimental limits for DY production of new physics are stronger. Therefore, due to the aforementioned, we will focus only on the DY type to set upper-bounds.

 It's worth mentioning that for both types of processes, we need to compute the effective cross section $\sigma_{eff}$, defined as
 \begin{equation}
     \sigma_{eff}=\epsilon \mathcal{A}\sigma,
 \end{equation}
 where $\epsilon$ stands for the detector efficiency and $\mathcal{A}$ represents the acceptance, defined as the ratio of the number of events satisfying the event selection criteria and the total number of selected events. A naive guess is to consider $\epsilon=0.55$, motivated by the efficiency of muon reconstruction \cite{muon_eff}, however, we consider two scenarios: $\epsilon=0.2$ and $\epsilon=0.55$ (The choice of the first value is well motivated by the results in Appendix \ref{sec:appendix}). On the other hand, the acceptance must be computed for each simulated sample, which must be done separately because the processes have different event selection criteria.

\begin{table}[!h]
    \centering
    \begin{tabular}{|c|c|c|c|c|}
    \hline
         & $M_{V^+}=200$[GeV]&$M_{V^+}=500$[GeV]&$M_{V^+}=800$[GeV]& ATLAS upper limit\\
        \hline
        $\sigma_{DY}$[fb] &$1.74\times 10^{-4}$& $1.32\times 10^{-6}$ & $6.56\times10^{-8}$ & $0.25$ \cite{susylims_dy}\\
         $\sigma_{VBF}$[fb] &   $7.86\times10^{-5}$    &   $2.06\times 10^{-7}$      &     $7.89\times 10^{-9}$   &  $2.67$ \cite{susylims}\\
         \hline
    \end{tabular}
    \caption{Benchmark values for the production cross section, for BP1 and a fixed value of $\beta_\mu=0.01$.}
    \label{bench_prel}
\end{table}
\section{Drell-Yan Production}\label{sec:dy}
\begin{figure}[!h]
\centering
    \begin{subfigure}{0.45\textwidth}
    \includegraphics[width=\textwidth]{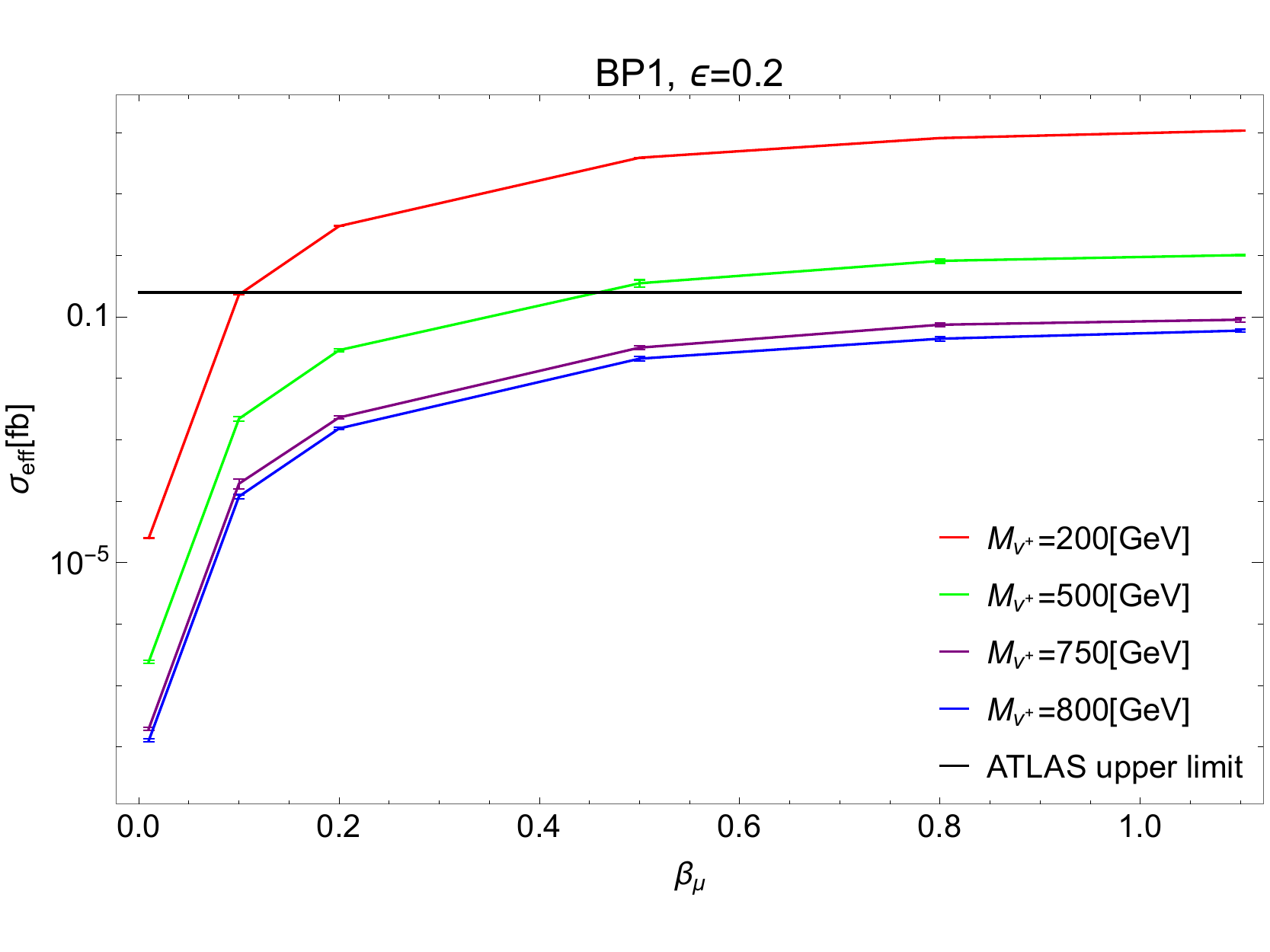}
    \caption{BP1, low $\beta_\mu$ values}
\end{subfigure}
\hspace{0.6cm}
\begin{subfigure}{0.45\textwidth}
    \centering
    \includegraphics[width=\textwidth]{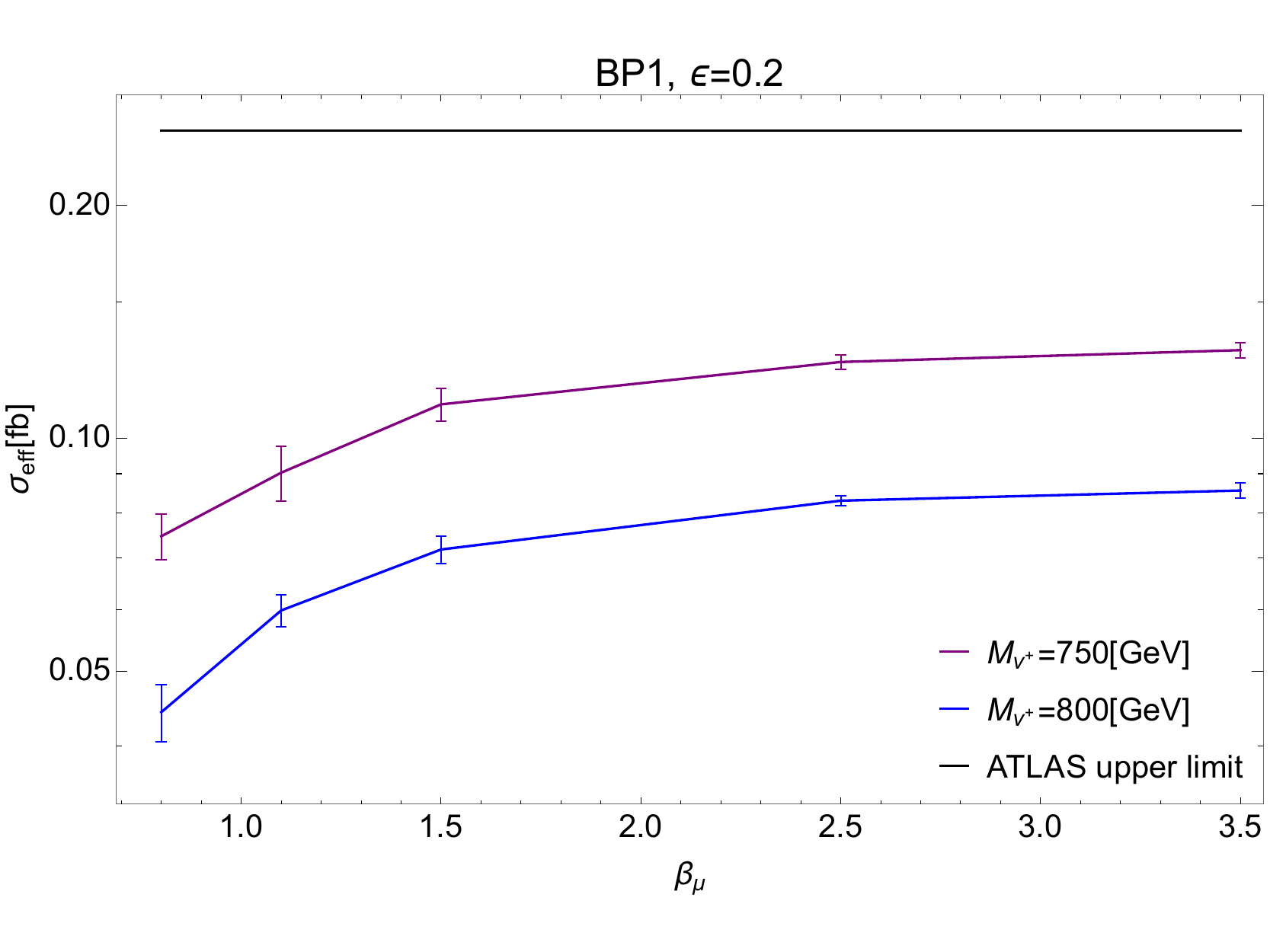}
    \caption{BP1, high $\beta_\mu$ values}
\end{subfigure}

    \begin{subfigure}{0.45\textwidth}
    \includegraphics[width=\textwidth]{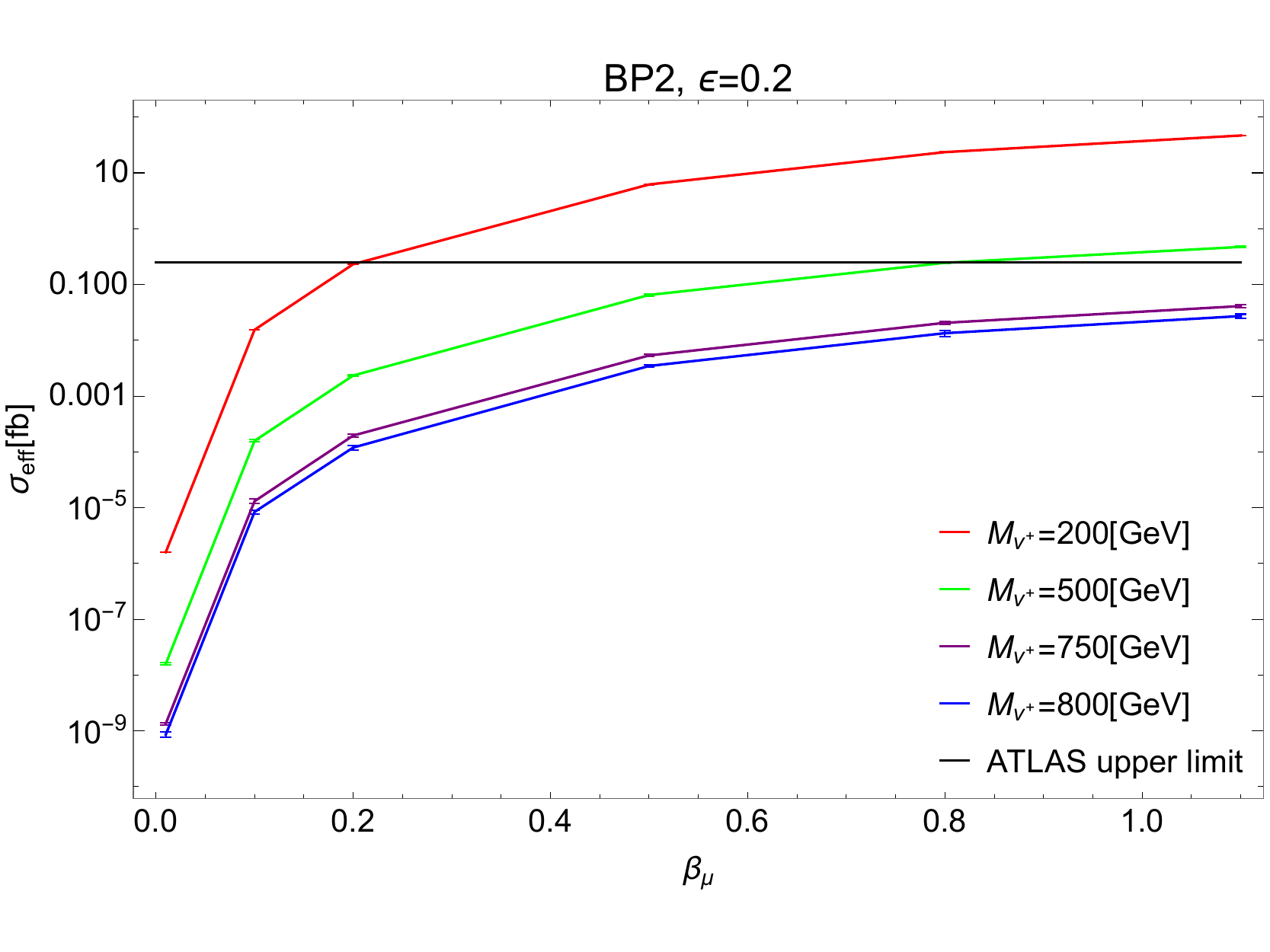}
    \caption{BP2, low $\beta_\mu$ values}
\end{subfigure}
\hspace{0.6cm}
\begin{subfigure}{0.45\textwidth}
    \centering
    \includegraphics[width=\textwidth]{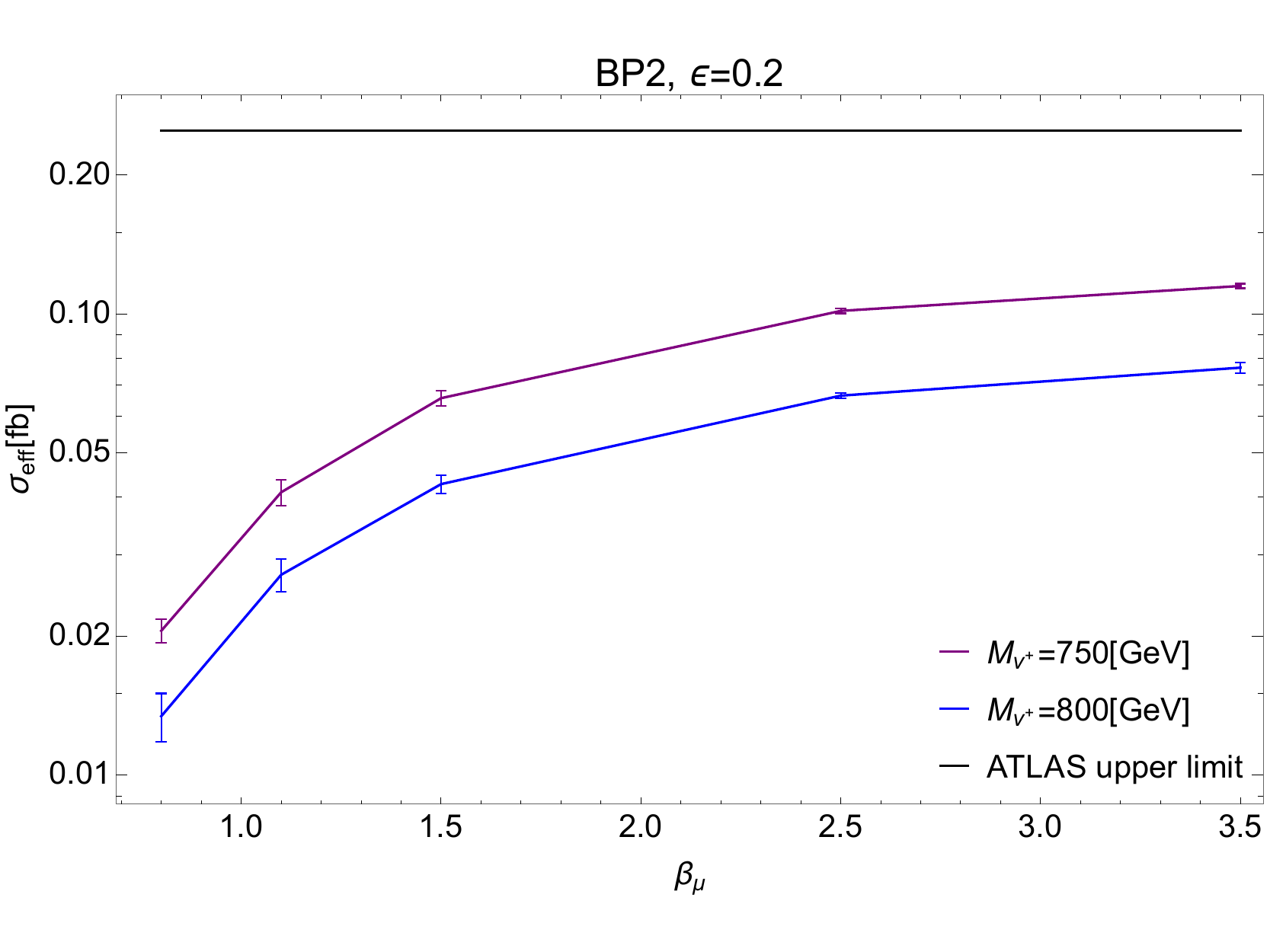}
    \caption{BP2, high $\beta_\mu$ values}
\end{subfigure}
\caption{Effective cross section for DY production, for $\epsilon=0.2$.}
\label{cross_dy_02}
\end{figure}

\begin{figure}[!h]
\centering
    \begin{subfigure}{0.45\textwidth}
    \includegraphics[width=\textwidth]{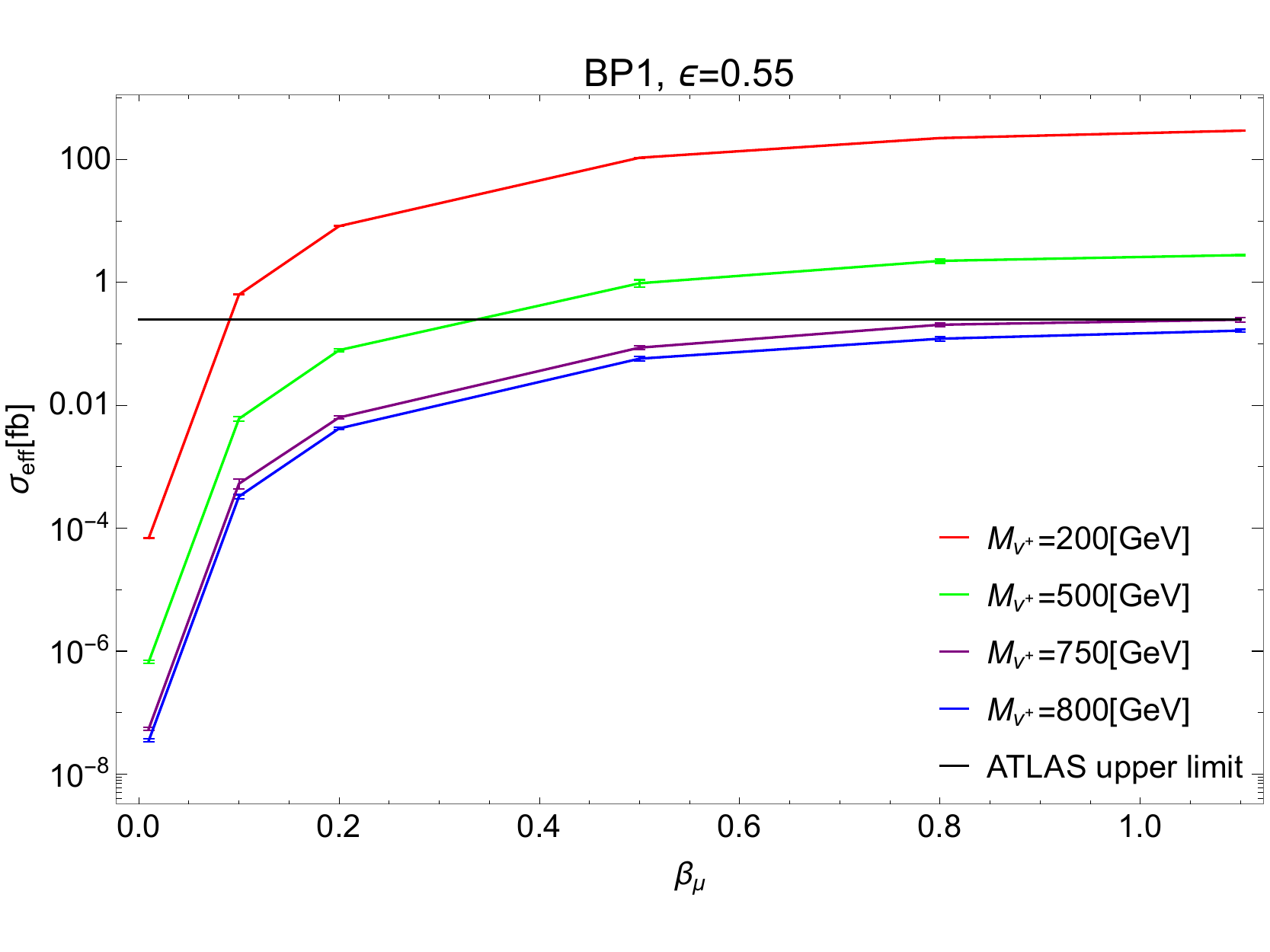}
    \caption{BP1, low $\beta_\mu$ values}
\end{subfigure}
\hspace{0.6cm}
\begin{subfigure}{0.45\textwidth}
    \centering
    \includegraphics[width=\textwidth]{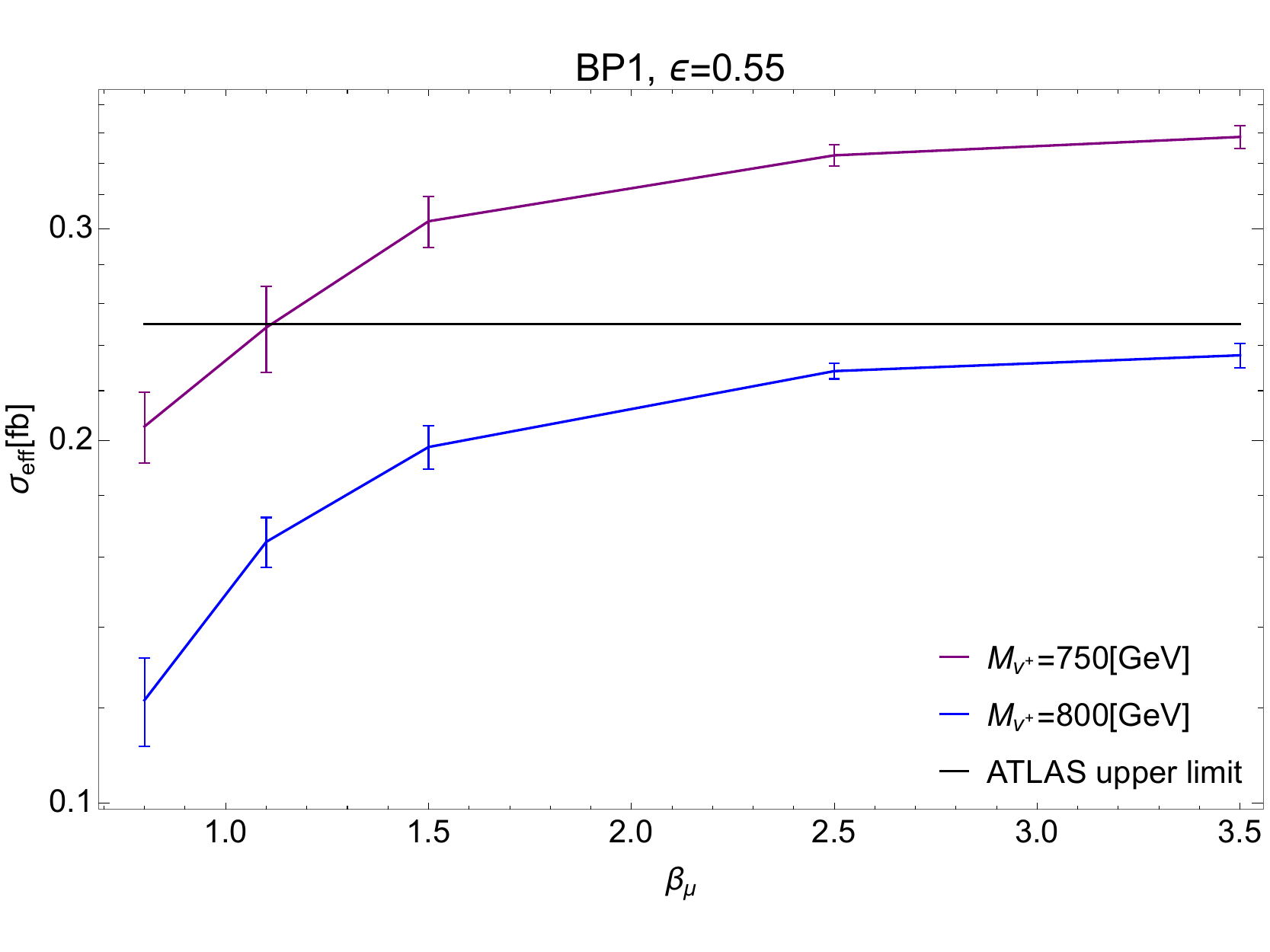}
    \caption{BP1, high $\beta_\mu$ values}
\end{subfigure}

    \begin{subfigure}{0.45\textwidth}
    \includegraphics[width=\textwidth]{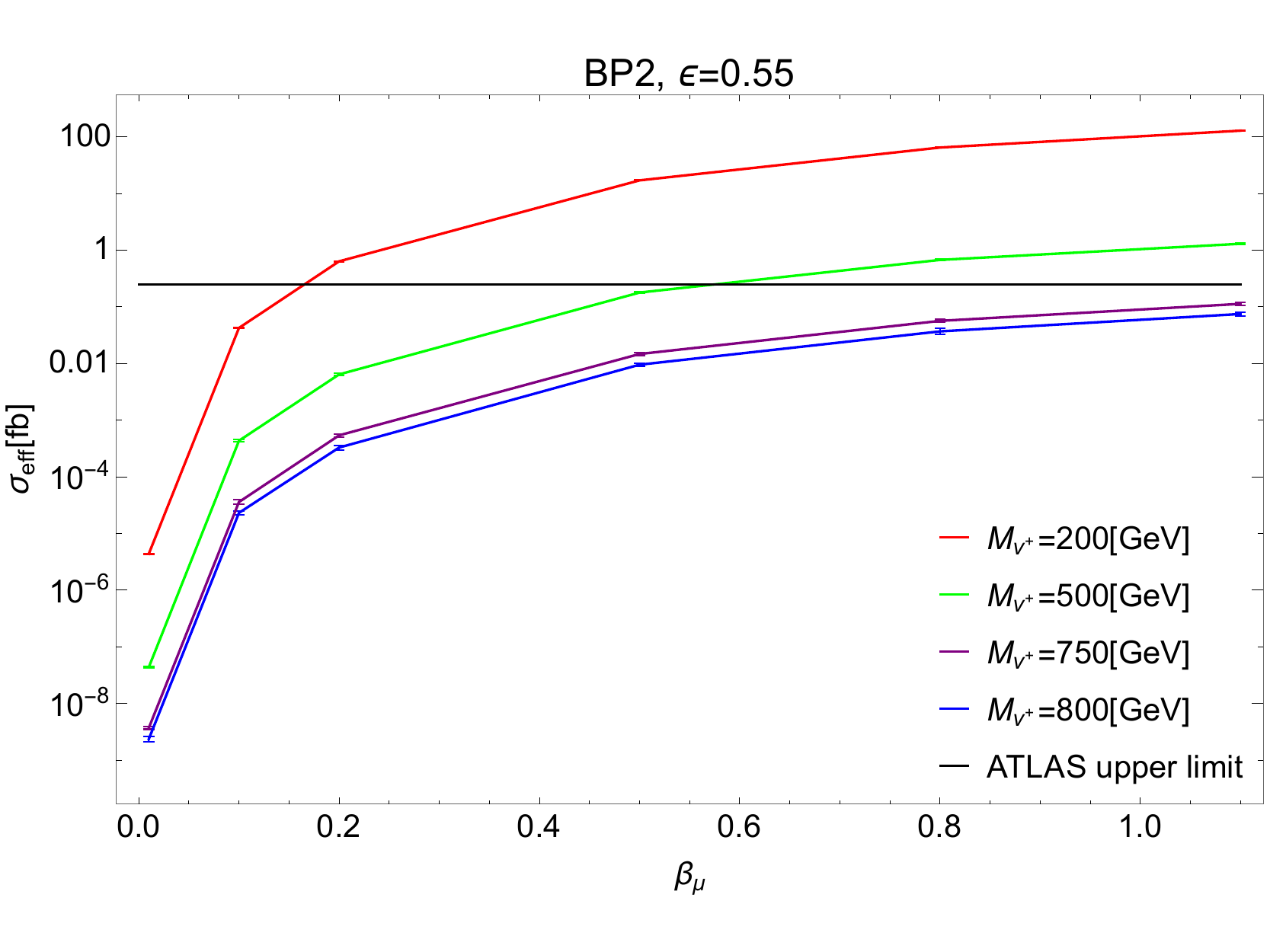}
    \caption{BP2, low $\beta_\mu$ values}
\end{subfigure}
\hspace{0.6cm}
\begin{subfigure}{0.45\textwidth}
    \centering
    \includegraphics[width=\textwidth]{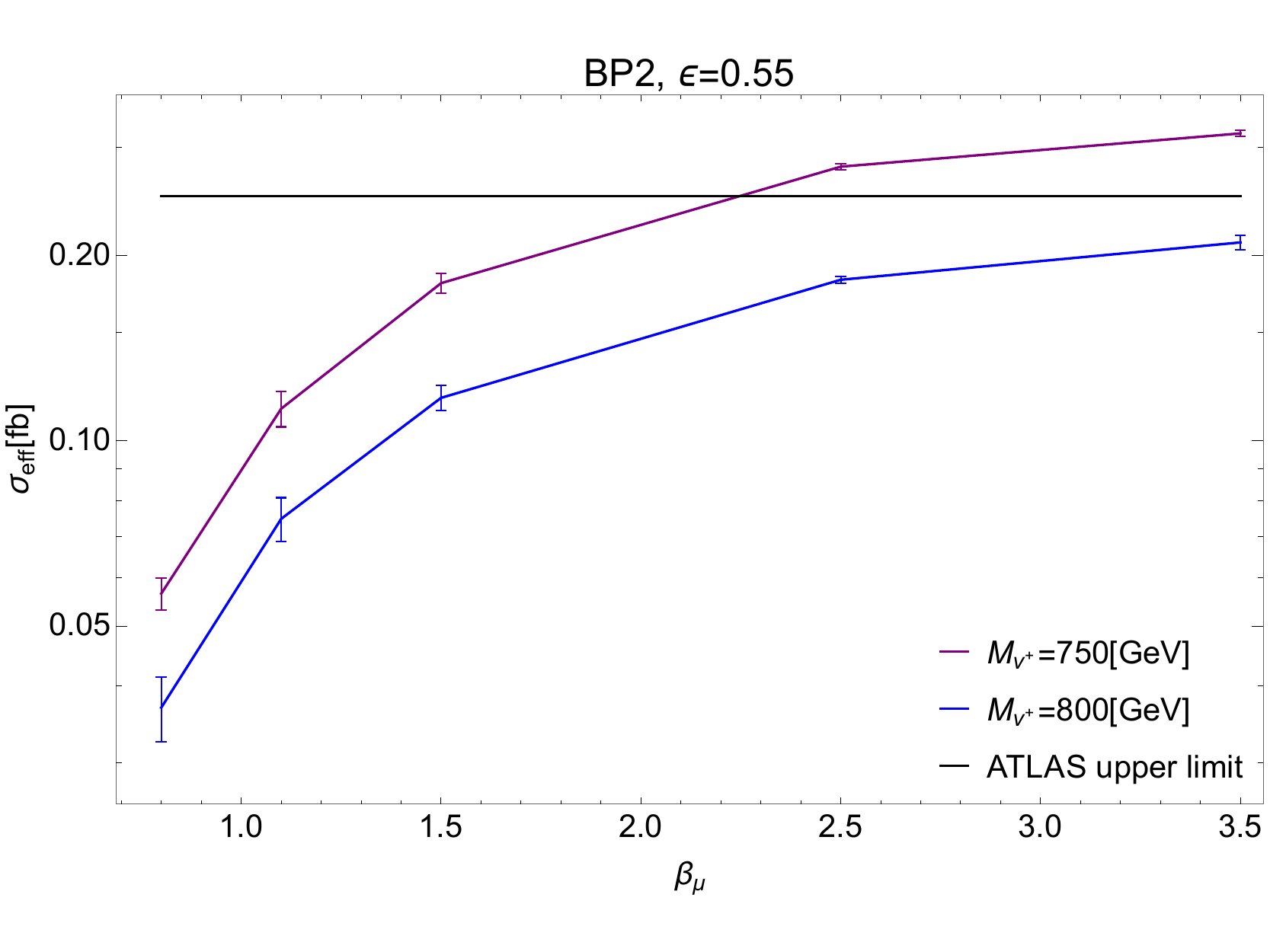}
    \caption{BP2, high $\beta_\mu$ values}
\end{subfigure}
\caption{Effective cross section for DY production, for $\epsilon=0.55$.}
\label{cross_dy_055}
\end{figure}

The current experimental limits for the DY production of new physics come from the search of sleptons at ATLAS \cite{susylims_dy}, in a final state composed of a lepton pair (electrons and muons) and missing energy. Since we have restricted our study to the case only with muons in the final state, the generated events must satisfy the selection criteria described in Table \ref{cut_def_dy}. As can be seen in Figures \ref{cross_dy_02} and \ref{cross_dy_055}, the limit on the $\beta_\mu$ coupling is more stringent in the low mass regime, and this limit relaxes as $M_{V^{+}}$ increases. Moreover, for $M_{V^{+}}=800$[GeV] the cross section is suppressed enough to escape the ATLAS upper limit, up to the perturbative scale ($\beta\sim \sqrt{4\pi}$). It's worth mentioning that these conclusions are true for both benchmark points, the main difference between them is that the values of the cross section are smaller for BP2. This suppression is explained by the contribution to the charged vector decay width, which is proportional to $|\beta_\tau|^2$.
On the other hand, we obtained the kinematical distributions for $M_{ll}$, $p_{T}^{miss}$, and $\cos \theta$, which is defined as the angle between the outgoing leptons. The model predicts longer tails for $M_{ll}$ and $p_{T}^{miss}$, as expected for SM extensions with heavy mediators, while the angular distribution presents a slightly distinctive pattern compared to the SM background.

\begin{table}[!h]
    \centering
    \begin{tabular}{|c|c|c|}
    \hline
     object    & definition &condition \\
     \hline
$p_T(l+)$ & Transverse momentum of the positively charged lepton & $> 25$ [GeV]\\
    $p_T(l-)$    & Transverse momentum of the negatively charged lepton & $> 25$[GeV]\\
    $\eta(l^+)$& pseudorapidity of the positively charged lepton & $<2.7$\\
        $\eta(l^-)$& pseudorapidity of the negatively charged lepton & $<2.7$\\
          $M_{ll}$ & Invariant mass of the SFOS lepton pair & $\geq 121.2$[GeV]\\
          $p_T^{miss}$ & Transverse component of the missing momentum vector &$>110$[GeV]\\
          \hline
    \end{tabular}
    \caption{Event selection criteria for DY production. These cuts were taken from reference \cite{susylims_dy}}
    \label{cut_def_dy}
\end{table}

\begin{figure}[!h]
    \begin{subfigure}{0.45\textwidth}
    \includegraphics[width=\textwidth]{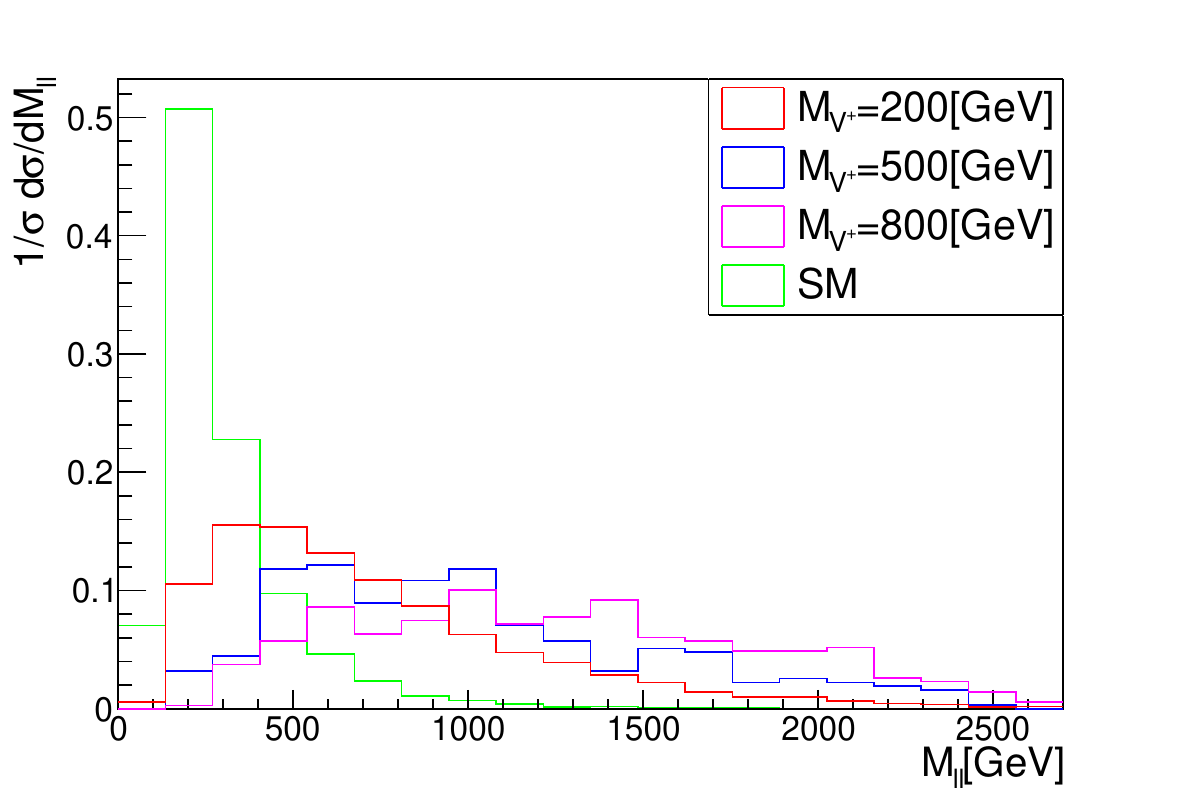}
    \caption{$M_{ll}$}
\end{subfigure}
    \begin{subfigure}{0.45\textwidth}
    \includegraphics[width=\textwidth]{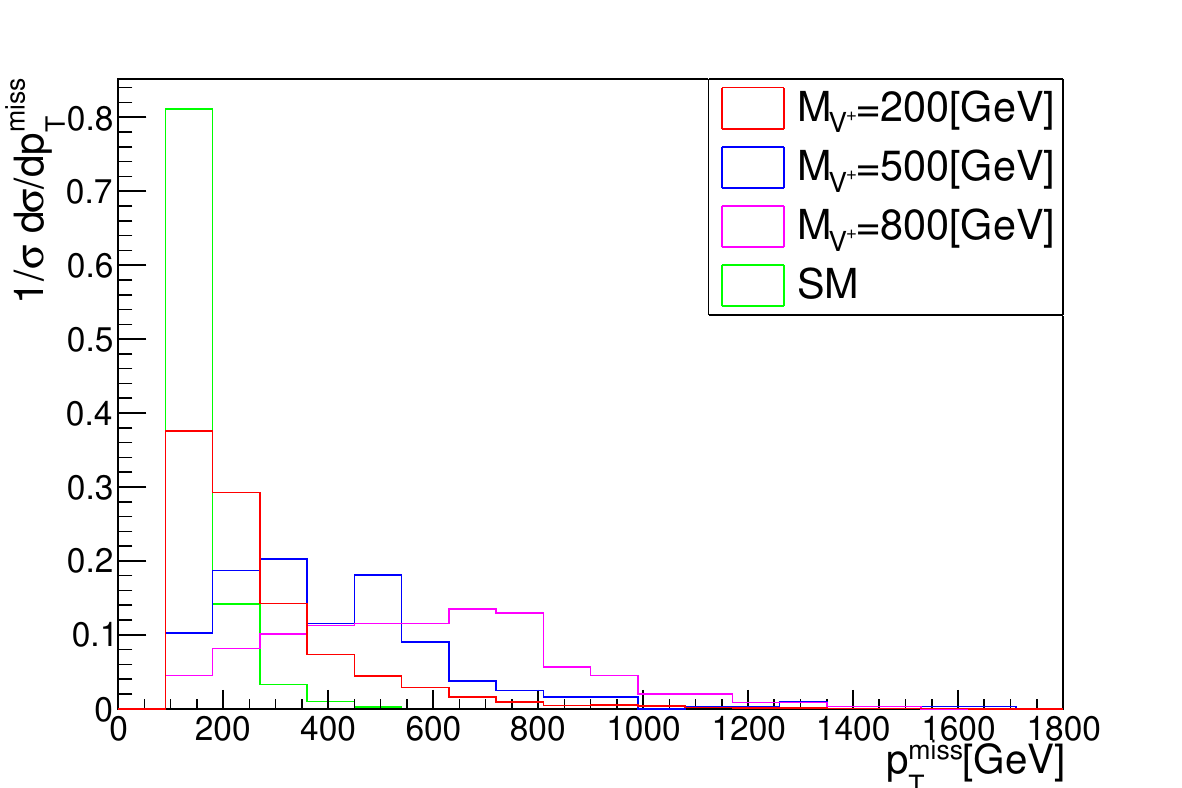}
    \caption{$p_T^{miss}$}
\end{subfigure}
\centering
    \begin{subfigure}{0.45\textwidth}
    \includegraphics[width=\textwidth]{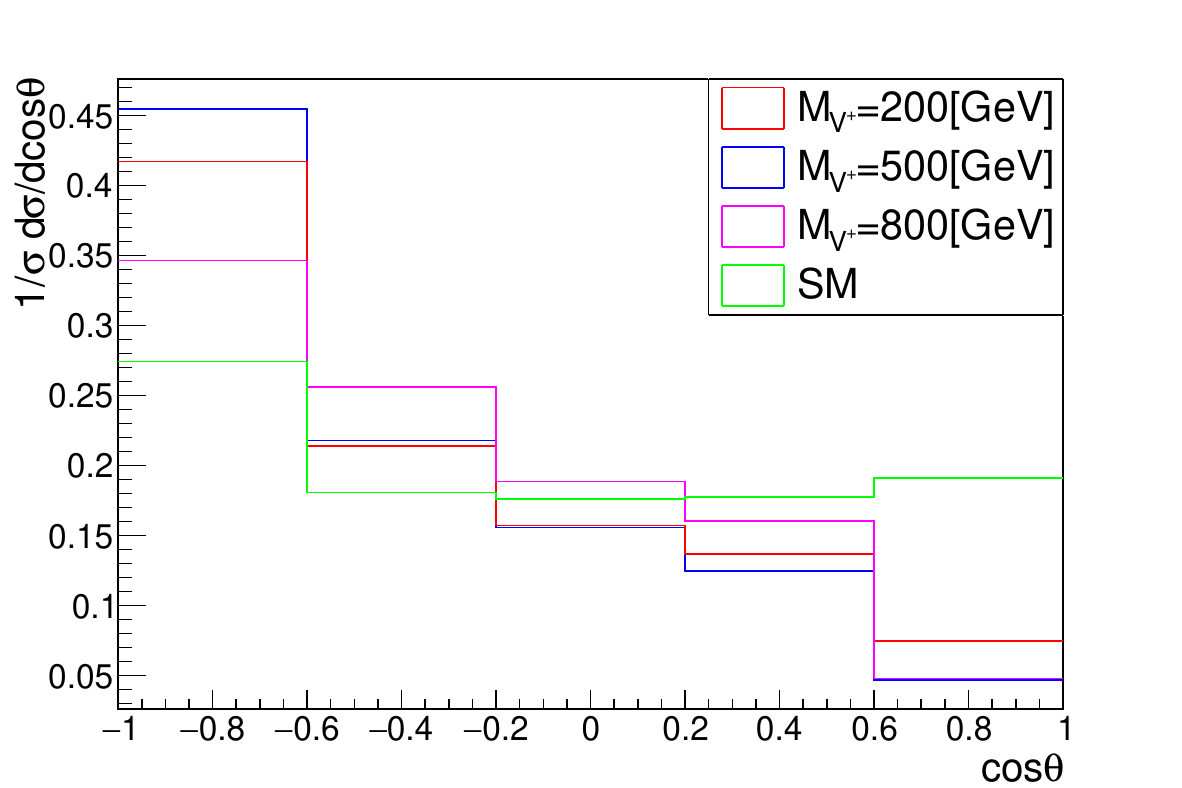}
    \caption{$\cos\theta$}
\end{subfigure}

\caption{Relevant kinematical distributions for DY production considering BP1 and $\beta_{\mu}=0.01$.}
\label{kin_dy}
\end{figure}

\section{Vector Boson Fusion production}\label{sec:vbf}
\begin{figure}[!h]
    \begin{subfigure}{0.45\textwidth}
    \includegraphics[width=\textwidth]{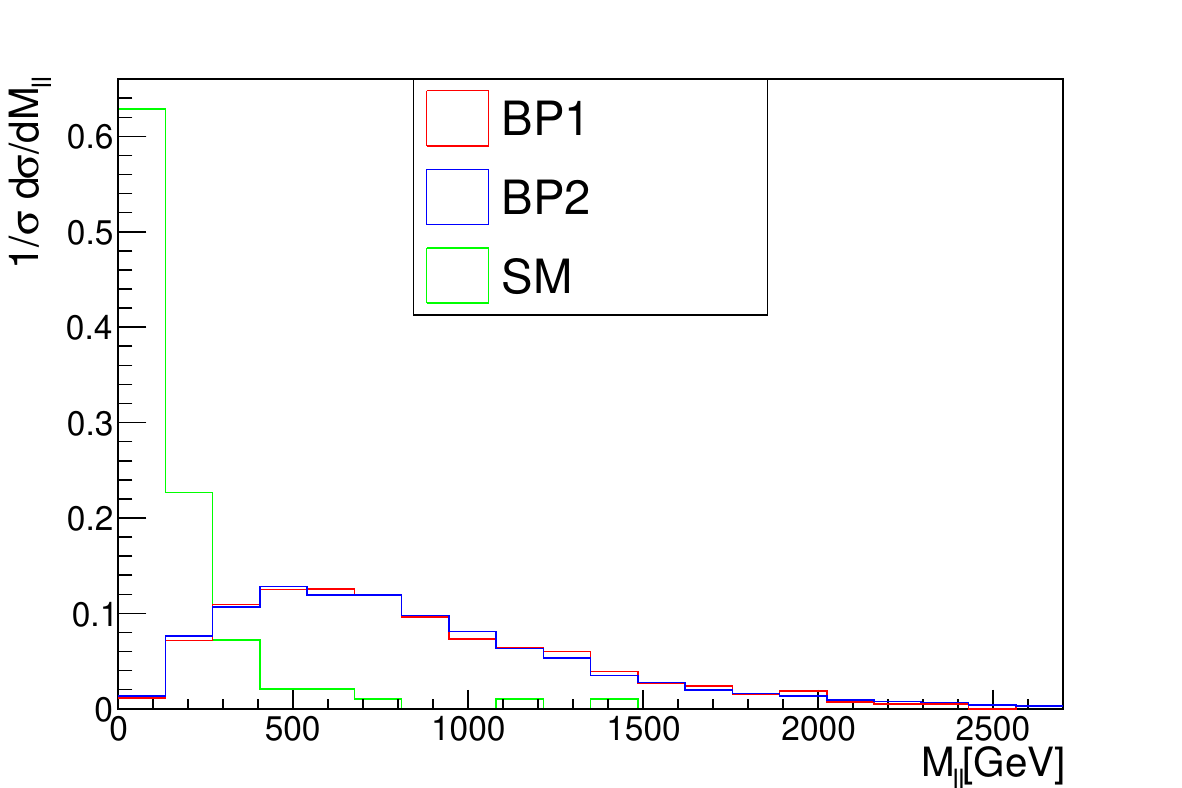}
    \caption{$M_{ll}$}
\end{subfigure}
    \begin{subfigure}{0.45\textwidth}
    \includegraphics[width=\textwidth]{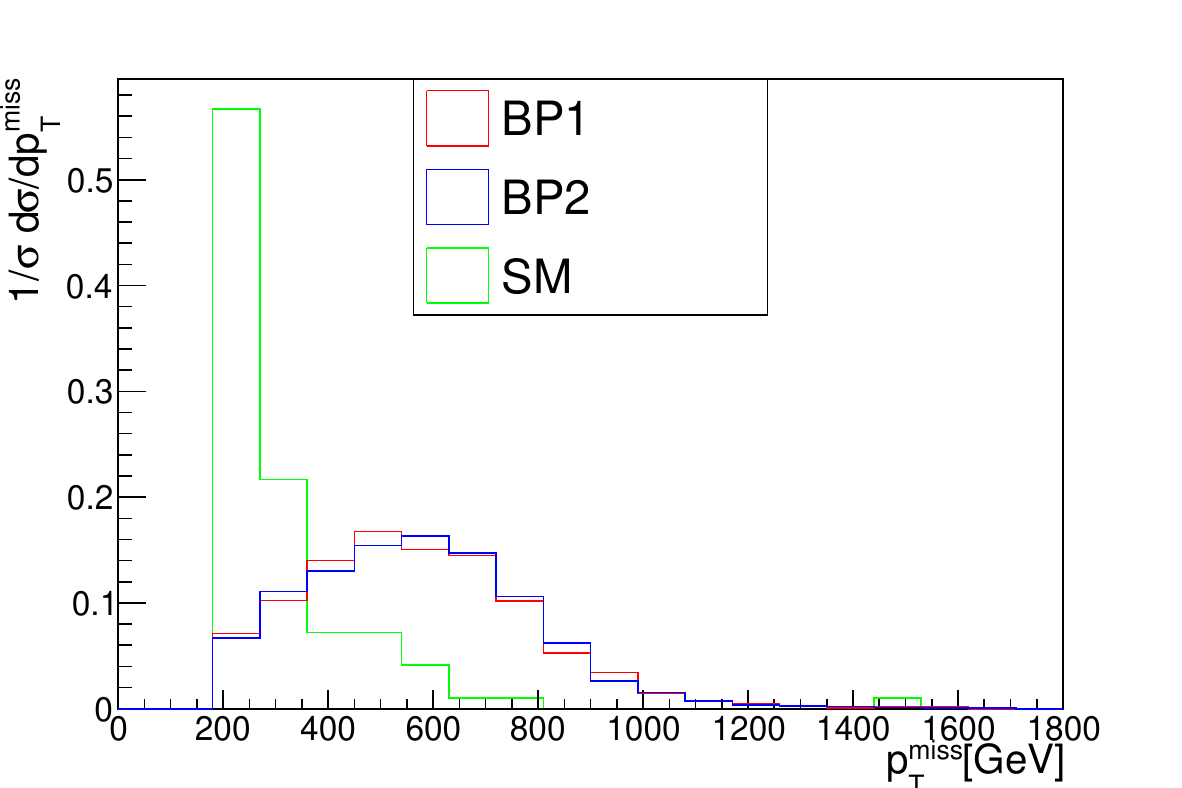}
    \caption{$p_T^{miss}$}
\end{subfigure}
    \begin{subfigure}{0.45\textwidth}
    \includegraphics[width=\textwidth]{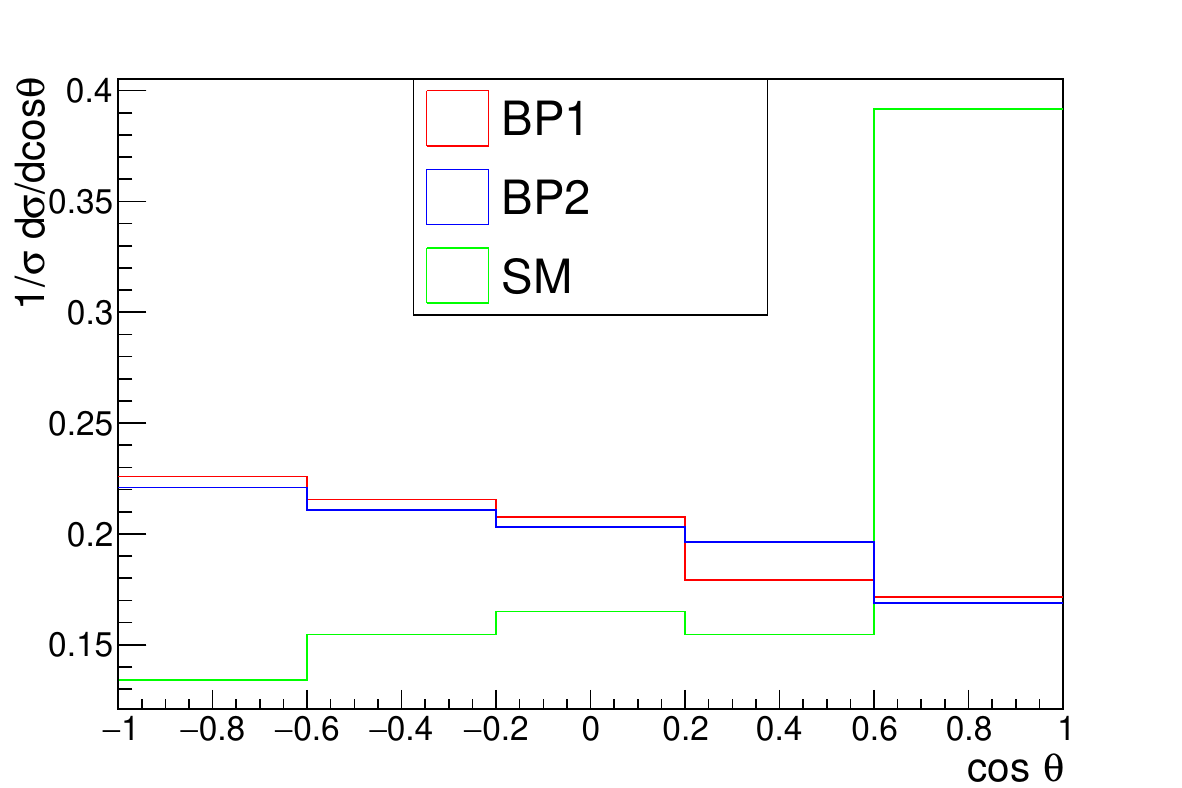}
    \caption{$\cos\theta$}
\end{subfigure}
    \begin{subfigure}{0.45\textwidth}
    \includegraphics[width=\textwidth]{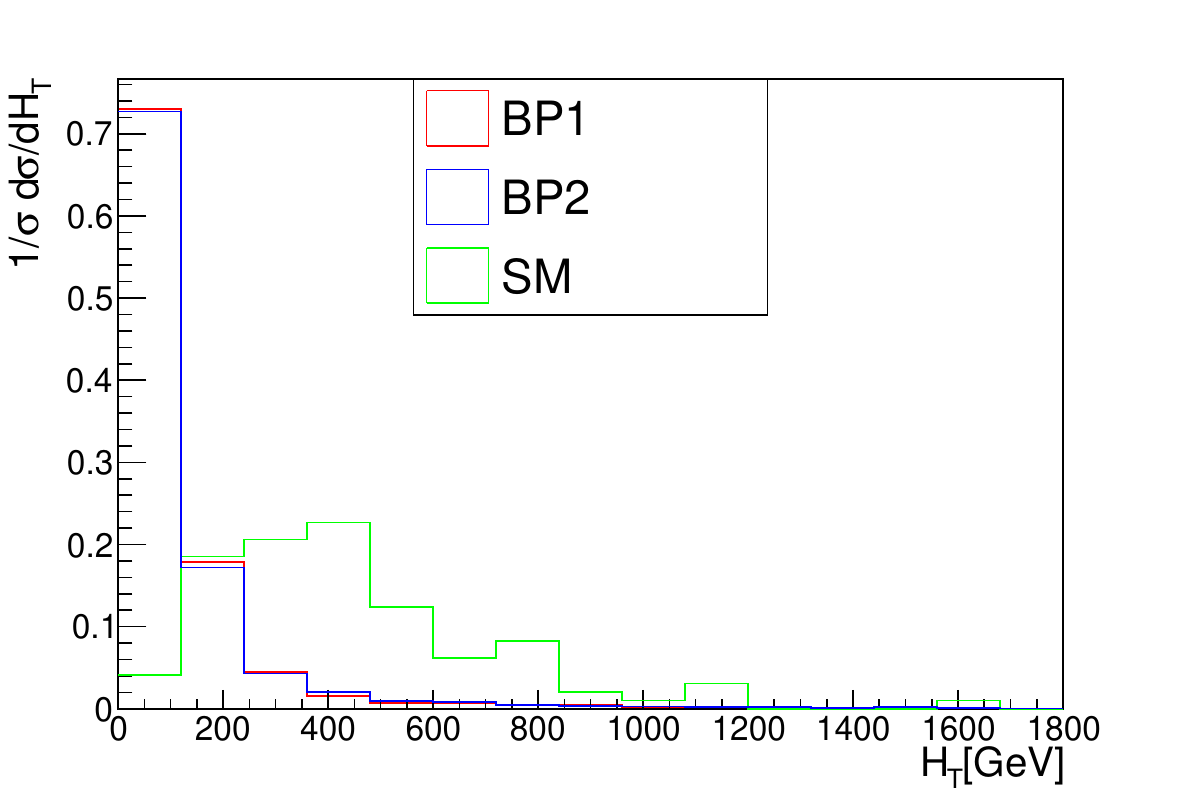}
    \caption{$H_T$}
\end{subfigure}

\caption{Relevant kinematical distributions for VBF production considering  $M_{V^{+}}=800$[GeV] and $\beta_{\mu}=3.5$.}
\label{kin_vbf}
\end{figure}

For VBF production, the event selection criteria are shown in Table \ref{cut_def_vbf}. Due to the strong limits arising from DY production, we have computed the VBF cross section for $M_{V^{+}}=800$[GeV] and $\beta_{\mu}=3.5$ for two benchmark points. Among the previously presented kinematical objects, we also considered the $H_T$ variable, defined as the scalar sum of the $p_T$ from all the outgoing jets. The effective cross sections can be seen in Table \ref{tab_cross_vbf}. While these cross sections are similar to the DY cross sections for the same parameter space points, the VBF background is significantly larger than the DY background. Indeed, taking the results from references \cite{susylims,susylims_dy}, we have found the following relation
\begin{equation}
    \frac{\sigma_{bkg}^{DY}}{\sigma_{bkg}^{VBF}}\approx 0.01.
\end{equation}
The big difference between the background cross sections makes the VBF significance considerably smaller under these conditions. The relevant kinematical distributions are depicted in Figure \ref{kin_vbf}. As can be seen, the results are similar to the DY kinematics, however, the angular distribution of the lepton pair in the background sample has less overlapping with the signal. This feature of the model can be used to implement a trigger, that can help to improve the discovery prospects under this production mechanism.

\begin{table}[!h]
    \centering
    \begin{tabular}{|c|c|c|c|c|}
    \hline
     & $\sigma_{eff} (\epsilon=0.2)$[fb]  & $\sigma_{eff}(\epsilon=0.55)$[fb]\\
     \hline
    BP1   &       $0.0502\pm 0.0011$       &         $0.138\pm 0.003$     \\
     BP2  &         $0.0487\pm 0.0007$     &   $0.134\pm 0.002$          \\
         \hline
    \end{tabular}
    \caption{Cross sections for VBF production, considering $\beta_\mu=3.5$ and $M_{V^+}=800$[GeV].}
    \label{tab_cross_vbf}
\end{table}

\begin{table}[!h]
    \centering
    \begin{tabular}{|c|c|c|}
    \hline
     object    & definition &condition \\
     \hline
$p_T(l+)$ & Transverse momentum of the positively charged lepton & $\geq 25$ [GeV]\\
    $p_T(l-)$    & Transverse momentum of the negatively charged lepton & $\geq 25$[GeV]\\
          $M_{ll}$ & Invariant mass of the SFOS lepton pair & $\geq 12$[GeV]\\
          $p_T^{miss}$ & Transverse component of the missing momentum vector &$>200$[GeV]\\
          $\Delta \phi(jet_{1},p_T^{miss})$ &Azimuthal separation between $p_T^{miss}$ and the first jet & $\geq 0.4$\\
            $\Delta \phi(jet_{2},p_T^{miss})$ &Azimuthal separation between $p_T^{miss}$ and the second jet & $\geq 0.4$\\
          \hline
    \end{tabular}
    \caption{Event selection criteria for VBF production. These cuts were taken from reference \cite{susylims}}
    \label{cut_def_vbf}
\end{table}

\section{Projection for future colliders}\label{sec:proj}
\begin{figure}[!h]
    \begin{subfigure}{0.45\textwidth}
    \includegraphics[width=\textwidth]{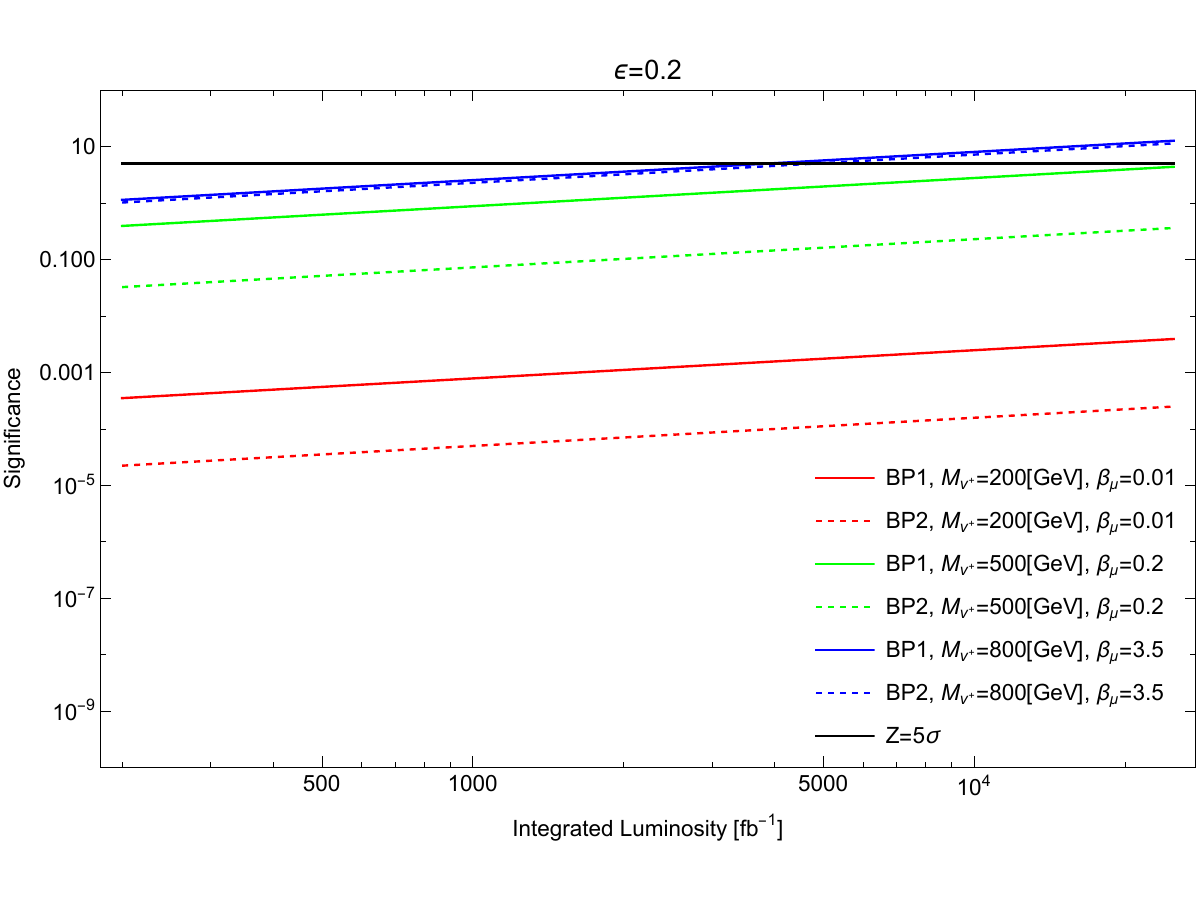}
    \caption{$\epsilon=0.2$}
\end{subfigure} \hspace{0.05\textwidth}
    \begin{subfigure}{0.45\textwidth}
    \includegraphics[width=\textwidth]{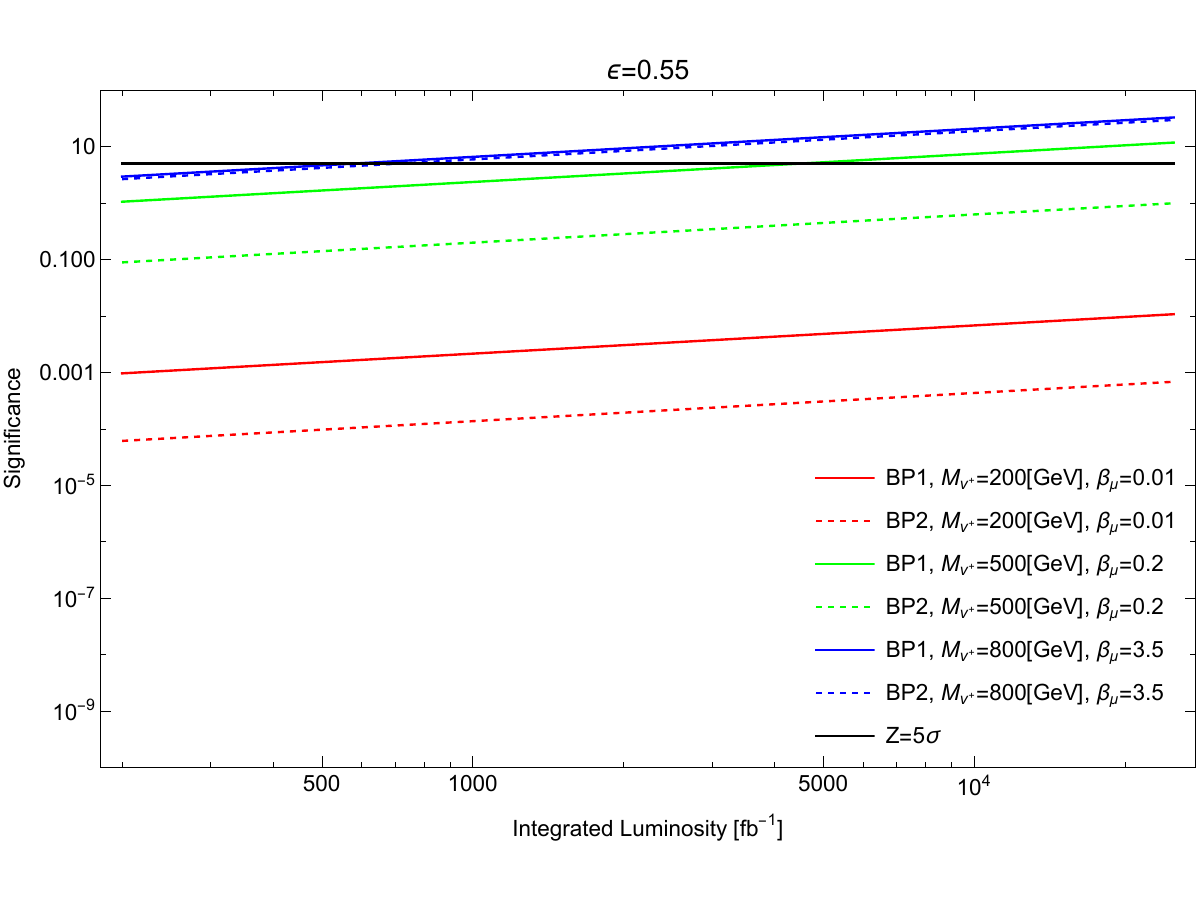}
    \caption{$\epsilon=0.55$}
\end{subfigure}

\caption{Significance projections for high luminosities.}
\label{proj_1}
\end{figure}

In order to study the discovery prospects of our model, we consider some scenarios that can be studied in future experiments. For this section, we restricted ourselves to DY production. Our predictions are based on the standard definition of significance ($Z$) 
\begin{equation}
    Z=\frac{s}{\sqrt{s+b}},
\end{equation}
where $s$ and $b$ stands for the number of signal and background events, respectively. Firstly, we have studied scenarios with fixed value of center of mass energy ($\sqrt{s}=13$[TeV]), this choice is motivated by the fact that the background cross section could be affected by $\sqrt{s}$. In Figure \ref{proj_1}, we show the $M_{V^+}=800$[GeV] scenario which is the most promising for luminosities that can be reached at early stage of the HL-LHC. On the other hand, the strong suppression on the couplings at the $M_{V^{+}}=200$[GeV] scenario makes it hard to probe in the near future, being practically undetectable during HL-LHC lifetime. It is worth to mention, that the lower mass regime ($M_{V^{+}}=200$[GeV]) could be probed in a scenario that consider the vector decay into tau leptons, however, this is not studied in this work. 

\begin{figure}[!h]
    \begin{subfigure}{0.45\textwidth}
    \includegraphics[width=\textwidth]{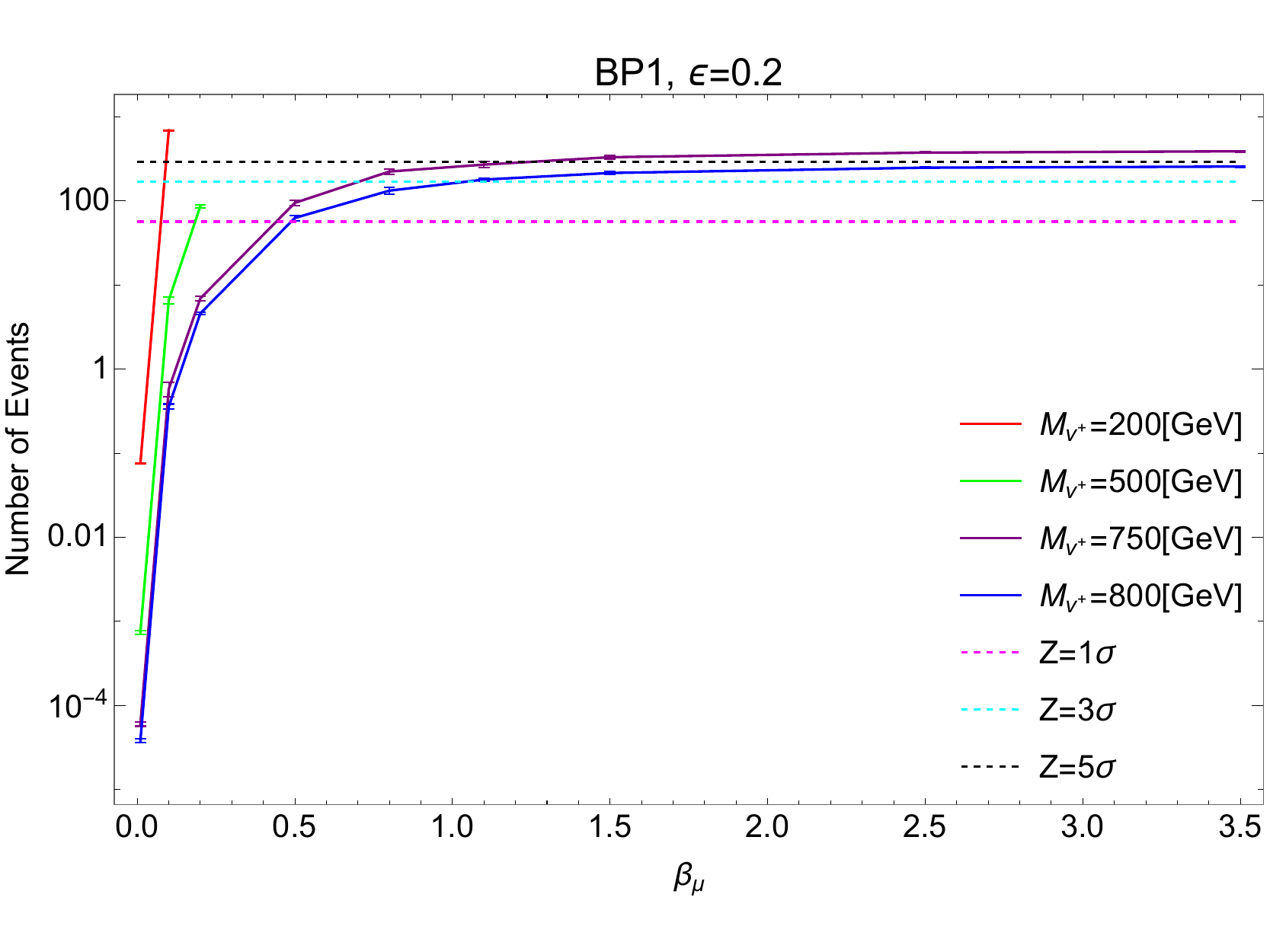}
    \caption{BP1, $\epsilon=0.2$}
\end{subfigure} \hspace{0.05\textwidth}
    \begin{subfigure}{0.45\textwidth}
    \includegraphics[width=\textwidth]{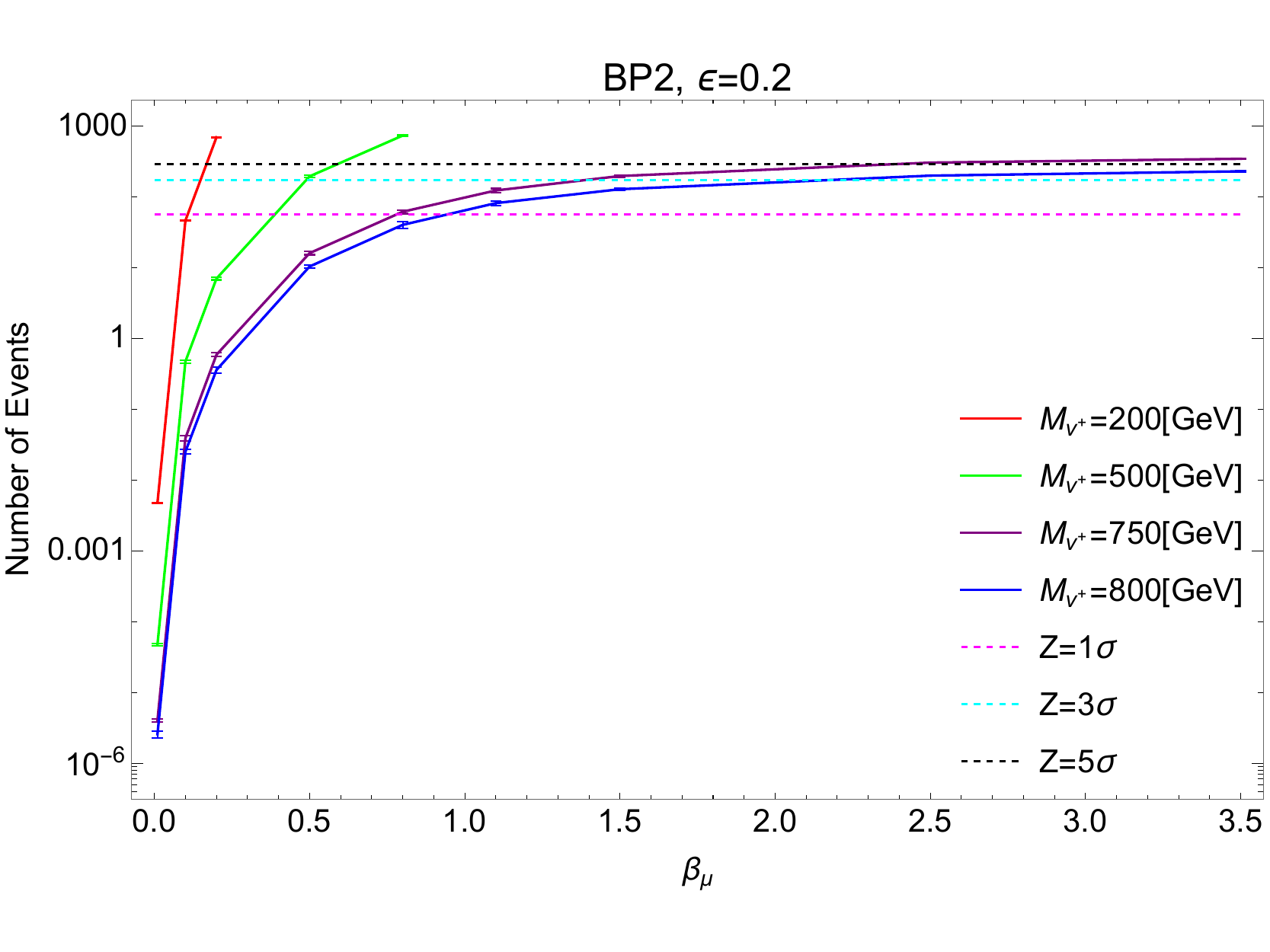}
    \caption{BP2, $\epsilon=0.2$}
\end{subfigure}
    \begin{subfigure}{0.45\textwidth}
    \includegraphics[width=\textwidth]{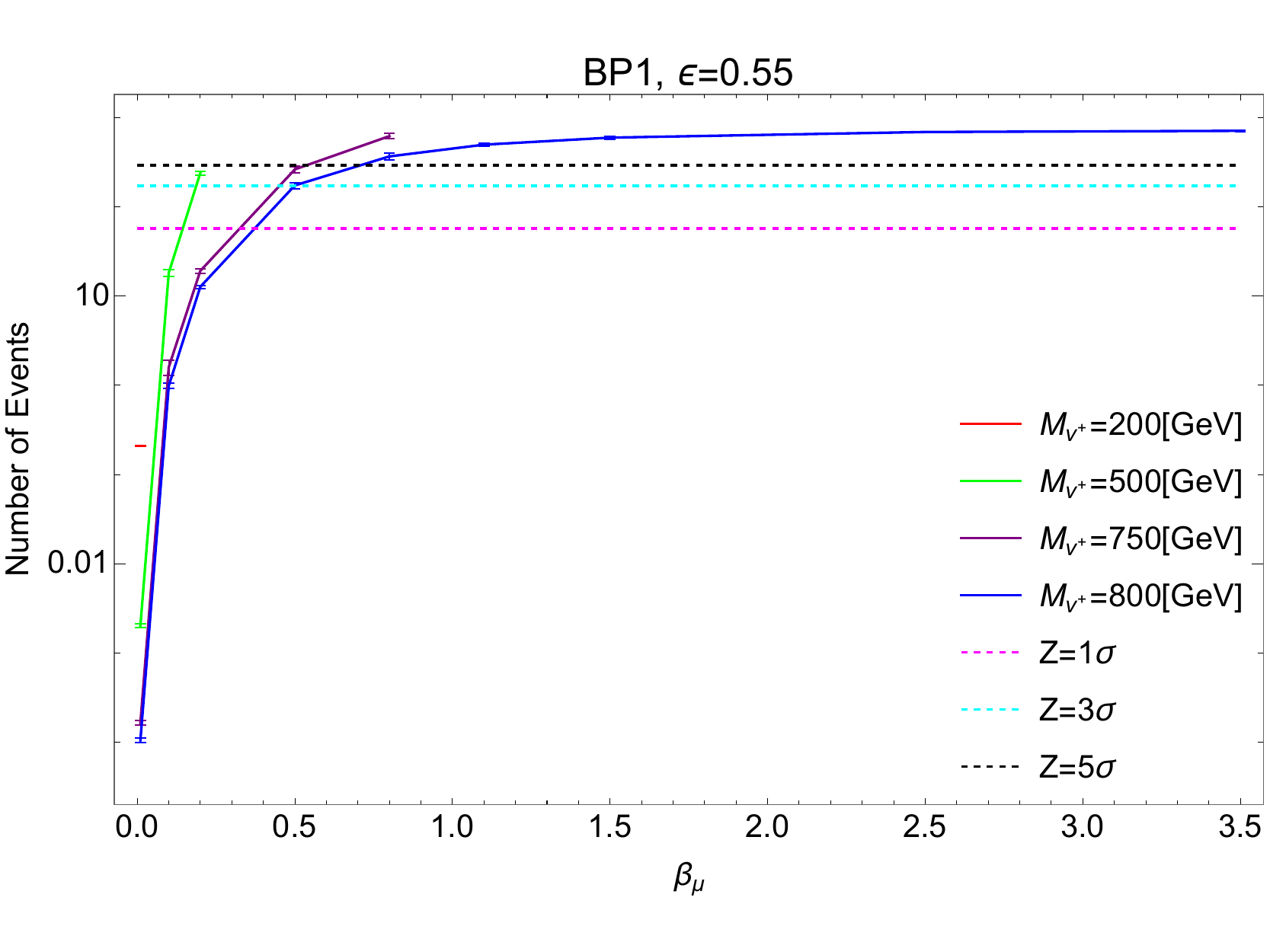}
    \caption{BP1, $\epsilon=0.55$}
\end{subfigure} \hspace{0.05\textwidth}
    \begin{subfigure}{0.45\textwidth}
    \includegraphics[width=\textwidth]{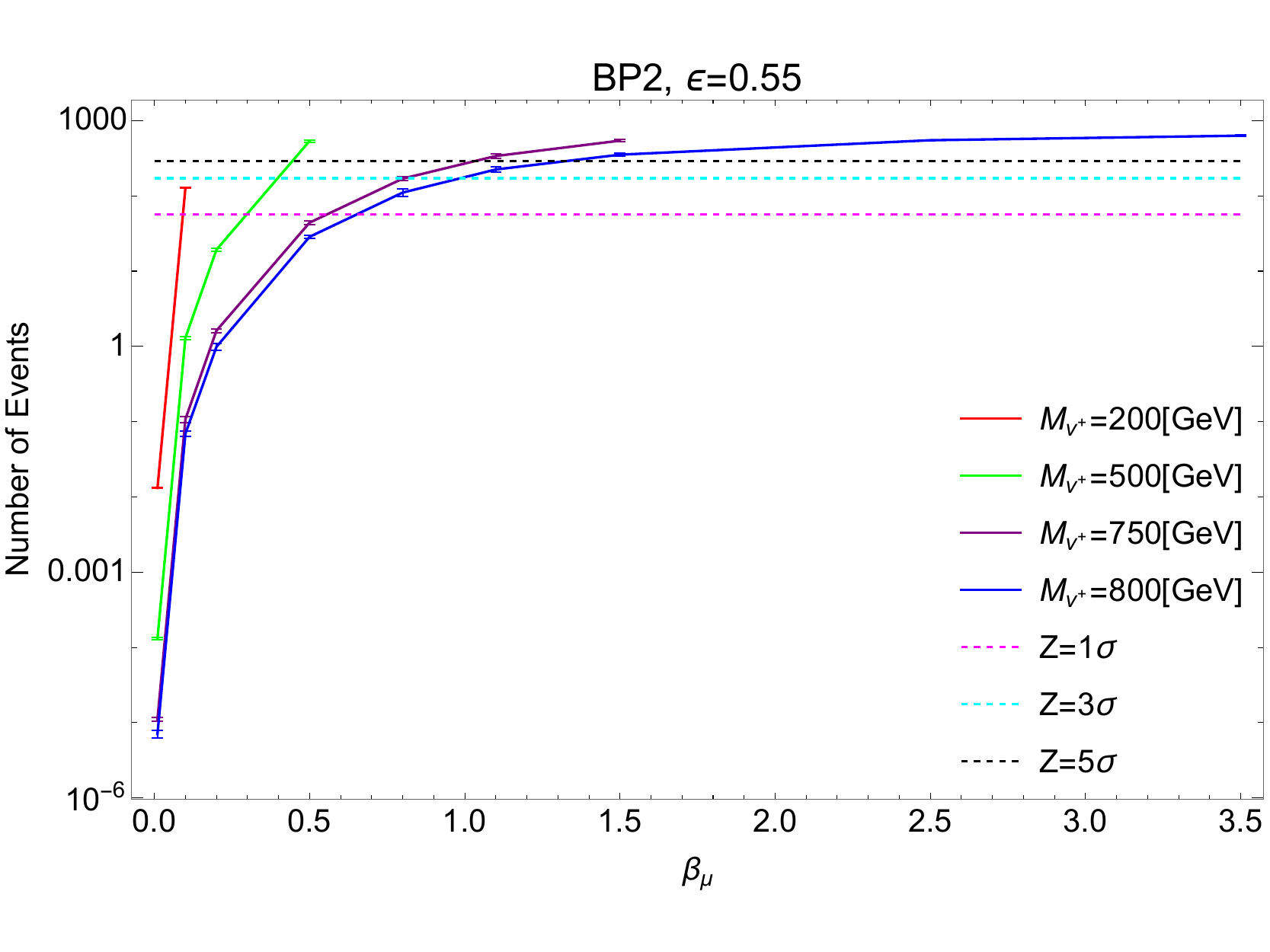}
    \caption{BP2, $\epsilon=0.55$}
\end{subfigure}

\caption{Expected number of events for  $\int L\mathrm{d}t=3000$[fb$^{-1}$] at $\sqrt{s}=13$[TeV]. The horizontal dotted lines stand for the number of events needed to overpass the corresponding significance level}
\label{proj_hilumi}
\end{figure}

In Figure \ref{proj_hilumi} we shows the expected number of event for $\int L\mathrm{d}t=3000$[fb$^{-1}$] at $\sqrt{s}=13$[TeV]. The horizontal dotted lines ($Z=1\sigma, 3\sigma, 5\sigma$) stand for the number of events needed to overpass certain significance level. Our result shows that a large region of the allowed parameter space, all the points above $Z=5\sigma$, can be probed during the HL-LHC lifetime

\section{Lower limits}\label{sec:lowlims}

Depending on the model kinematical regime, some constraints on the couplings of the HNL can be obtained from astrophysical observations. For instance, when the HNL is a DM candidate, we can use the DM abundance measured by Planck \cite{PLANCK} as input to set limits.

According to reference \cite{dong2021} the annihilation cross section for the left handed HNL has the following form
\begin{equation}
    \langle \sigma v\rangle=\sum_{k,k'=\{e,\mu,\tau\}}|\beta_k^*\beta_{k'}|^2\frac{M_N^2}{8\pi}\left(1+\frac{8T_f}{M_N}\right)\left(\frac{1}{M_{V^+}^4}+\frac{4}{(M_{V^1}^2+M_{V^2}^2)^2}\right).
\end{equation}

\begin{figure}
    \centering
    \includegraphics[width=0.5\textwidth]{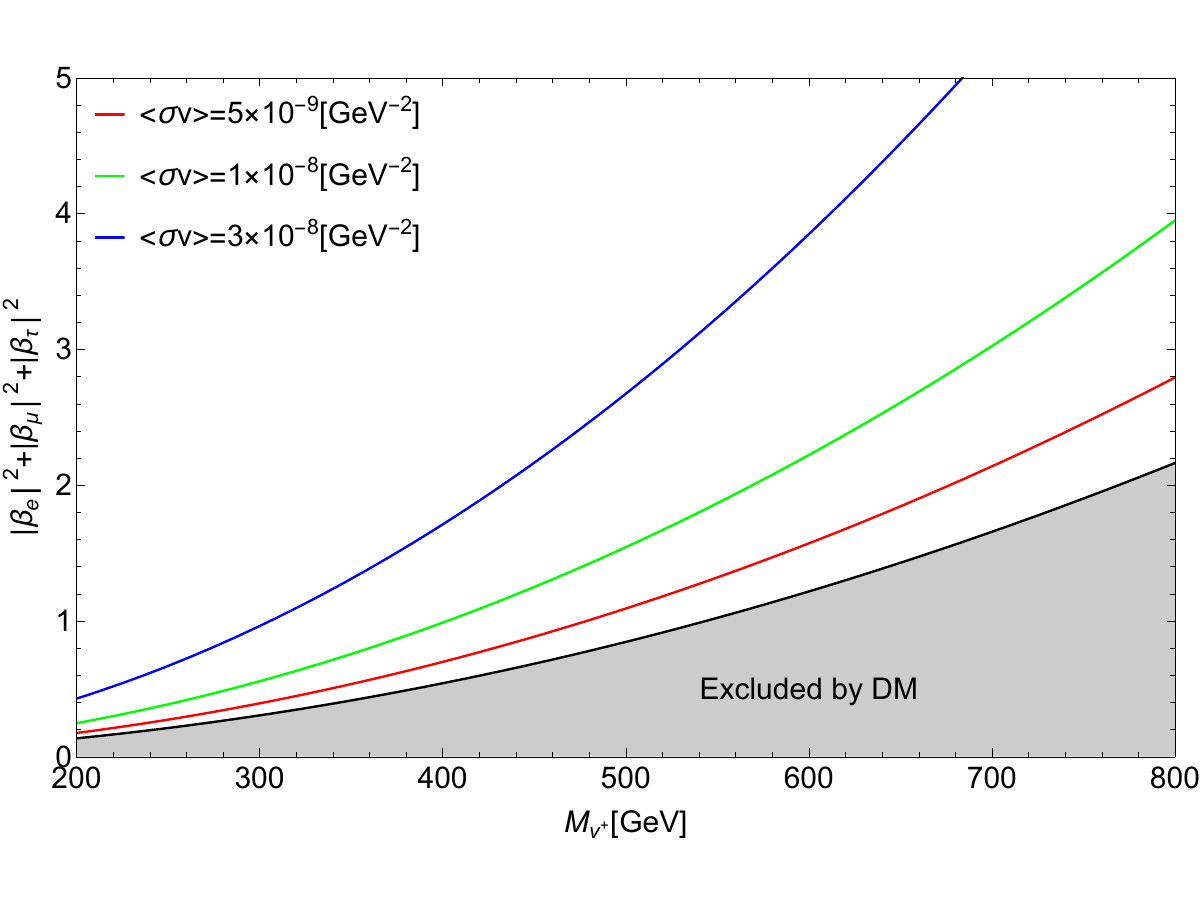}
    \caption{Allowed parameter space considering the DM constraint.}
    \label{dmlims}
\end{figure}

In order to avoid overabundance, the thermally averaged cross section must satisfy the following lower bound (for instance, see references \cite{langacker,pdg2022,Profumo}):
\begin{equation}
    \langle \sigma v\rangle \geq 3\times10^{-9}[\text{GeV}^{-2}],
\end{equation}

which provides a restriction for the parameter space as seen in Figure \ref{dmlims}.

\begin{figure}[!h]
    \begin{subfigure}{0.45\textwidth}
    \includegraphics[width=\textwidth]{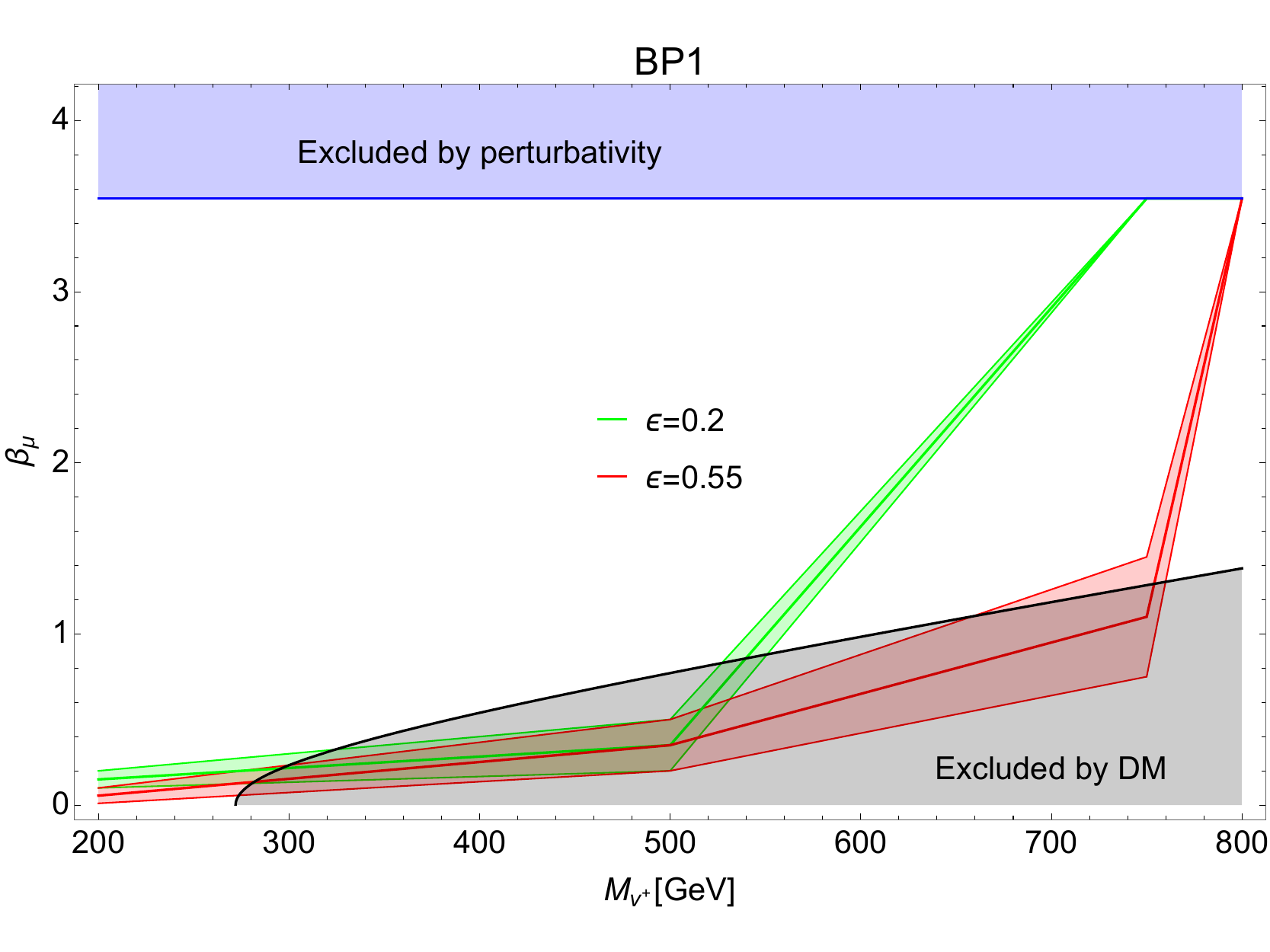}
    \caption{BP1}
\end{subfigure}\hspace{0.05\textwidth}
    \begin{subfigure}{0.45\textwidth}
    \includegraphics[width=\textwidth]{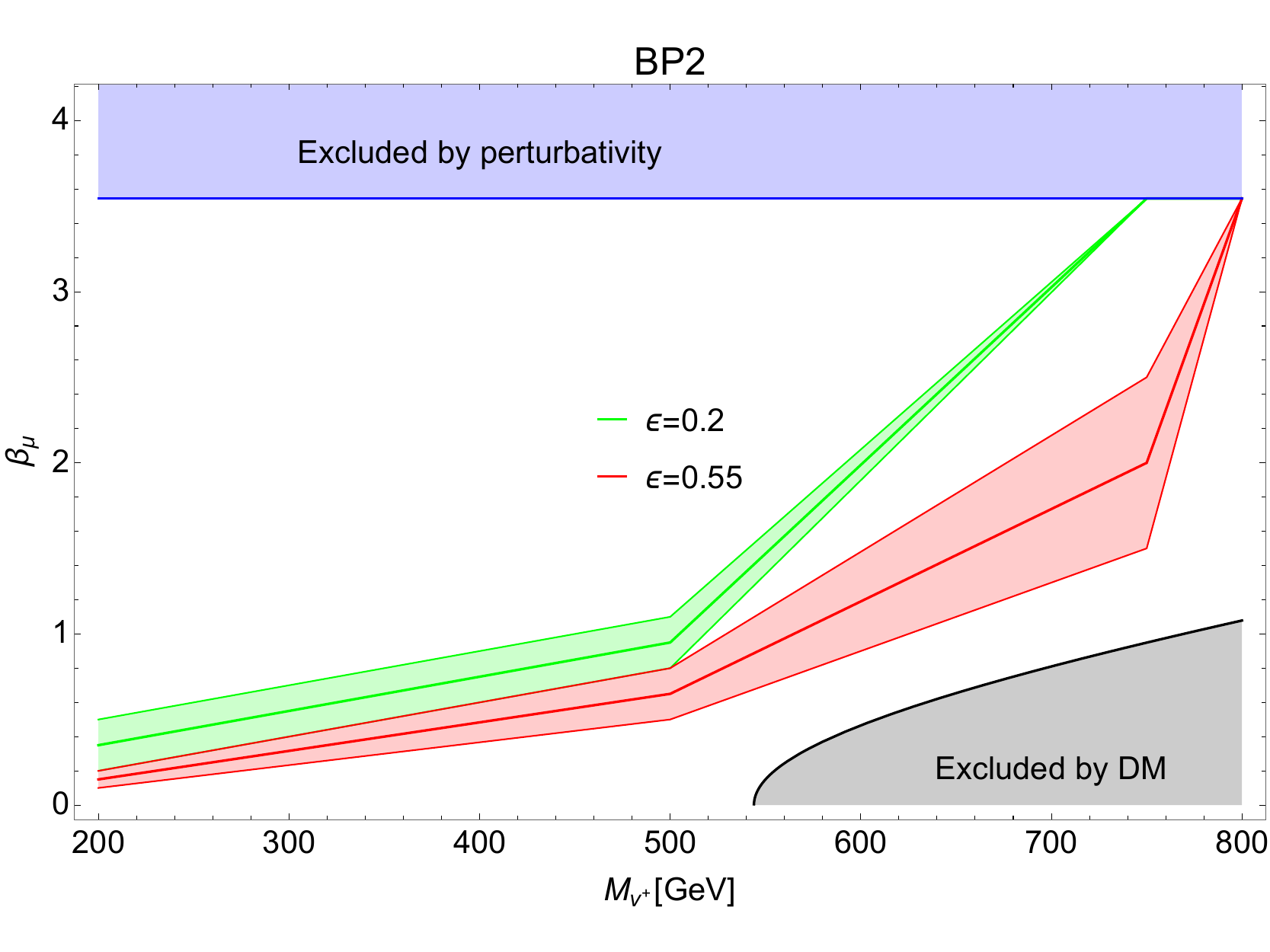}
    \caption{BP2}
\end{subfigure}

\caption{Combined limits on the parameter space.}
\label{comblims}
\end{figure}

The collider limits presented in this work can be combined with the DM limits, as is shown in Figure \ref{comblims}, there is an interesting interplay between the upper limits from ATLAS and the lower limit from dark matter. Due to this interplay, a large part of the parameter space in BP1 seems to be excluded by dark matter abundance, however, it's worth emphasizing that the lower limit is a rough approximation and a rigorous study on dark matter relic abundance can help to precisely rule out that part of the parameter space (the grey region in Figure \ref{comblims}).

\section{Conclusions}\label{sec:conc}
In this work, we have studied a simplified extension of the SM based on a Massive Vector Doublet and a Left-Handed HNL. The results have shown that the model produces a characteristic signal composed by a SFOS lepton pair and missing energy. While this state has been studied in the past, our analysis shows that the angular distribution of the lepton pair is a key variable for the identification of new physics in this final state. 
Indeed, it may constitute a discovery method in future experiments. Moreover, the kinematical analysis presented in this work could be repeated for different BSM scenarios related to this final state, and the angular distributions could be useful to classify different models.
We have performed an analysis to set limits on the parameter space using experimental cuts applied by current searches of new physics and considering two different prescriptions for the detector efficiency:  $\epsilon=0.2$ and $\epsilon=0.55$. The first is well motivated by the simulation of the detector response and the latter is an optimistic prediction considering future developments of the LHC. The main effect of the efficiency is reflected in the definition of upper limits on the parameter space, these limits get relaxed for the lower efficiency. However, the remaining parameter space can be probed at the HL-LHC in both prescriptions.
\newline
On the other hand, we have constrained the parameter space based on the dark matter relic abundance. The lower bound depends strongly on the exotic neutrino coupling to the three families of leptons, however, we restricted our analysis only to the coupling to muons. One possible way to constrain all flavors is to consider  Lepton Flavor Violation (LFV) processes. These type of processes can be suited in the model at 1 loop, but the calculation of these loops is not trivial, due to the non minimal gauge couplings. Therefore, a rigorous study of the radiative processes will be carried out in a future work.



\appendix

\section{Analysis validation}\label{sec:appendix}
\begin{figure}[!h]
\centering
    \begin{subfigure}{0.45\textwidth}
    \includegraphics[width=\textwidth]{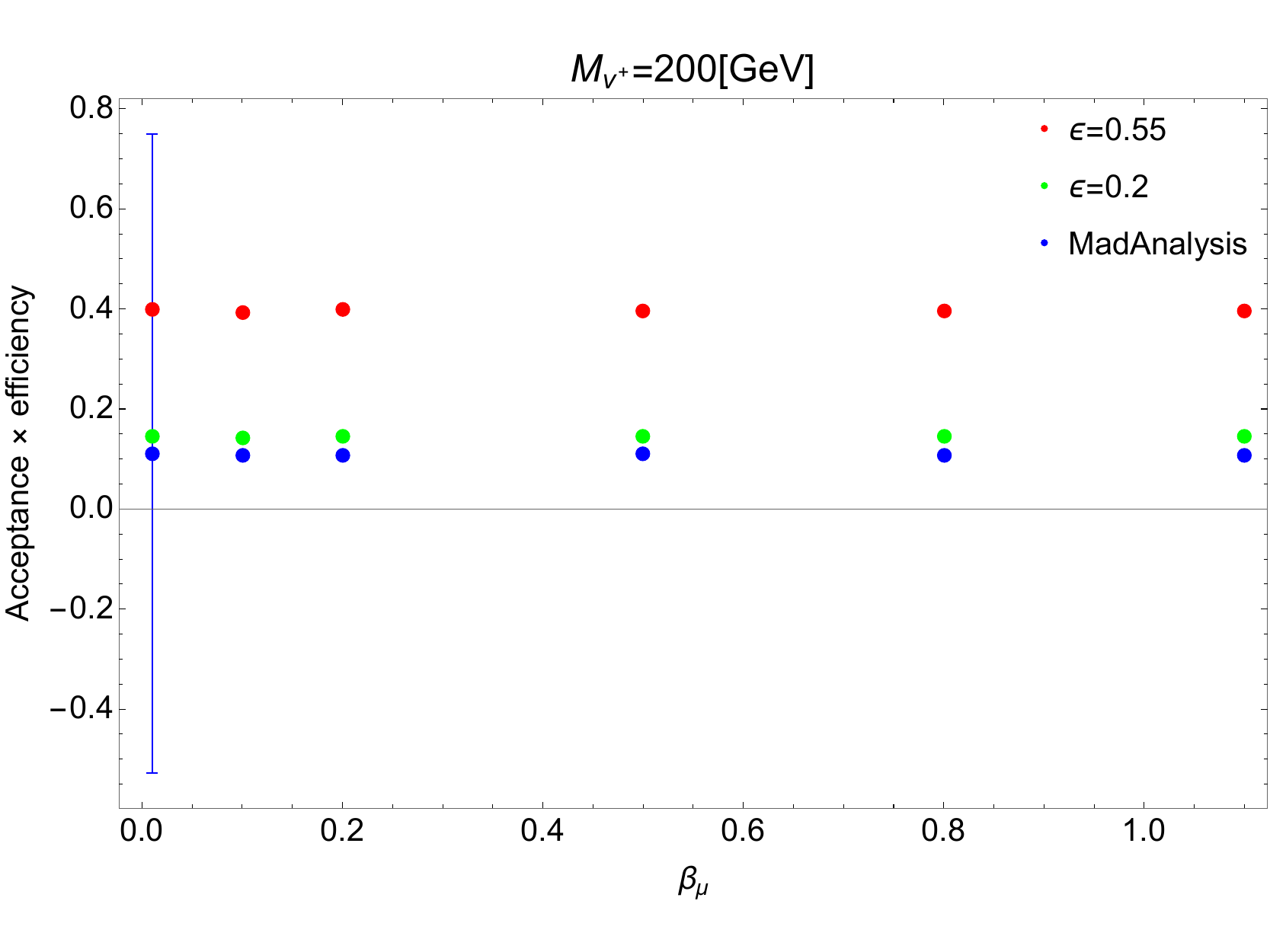}
    \caption{}
\end{subfigure}
\hspace{0.6cm}
\begin{subfigure}{0.45\textwidth}
    \centering
    \includegraphics[width=\textwidth]{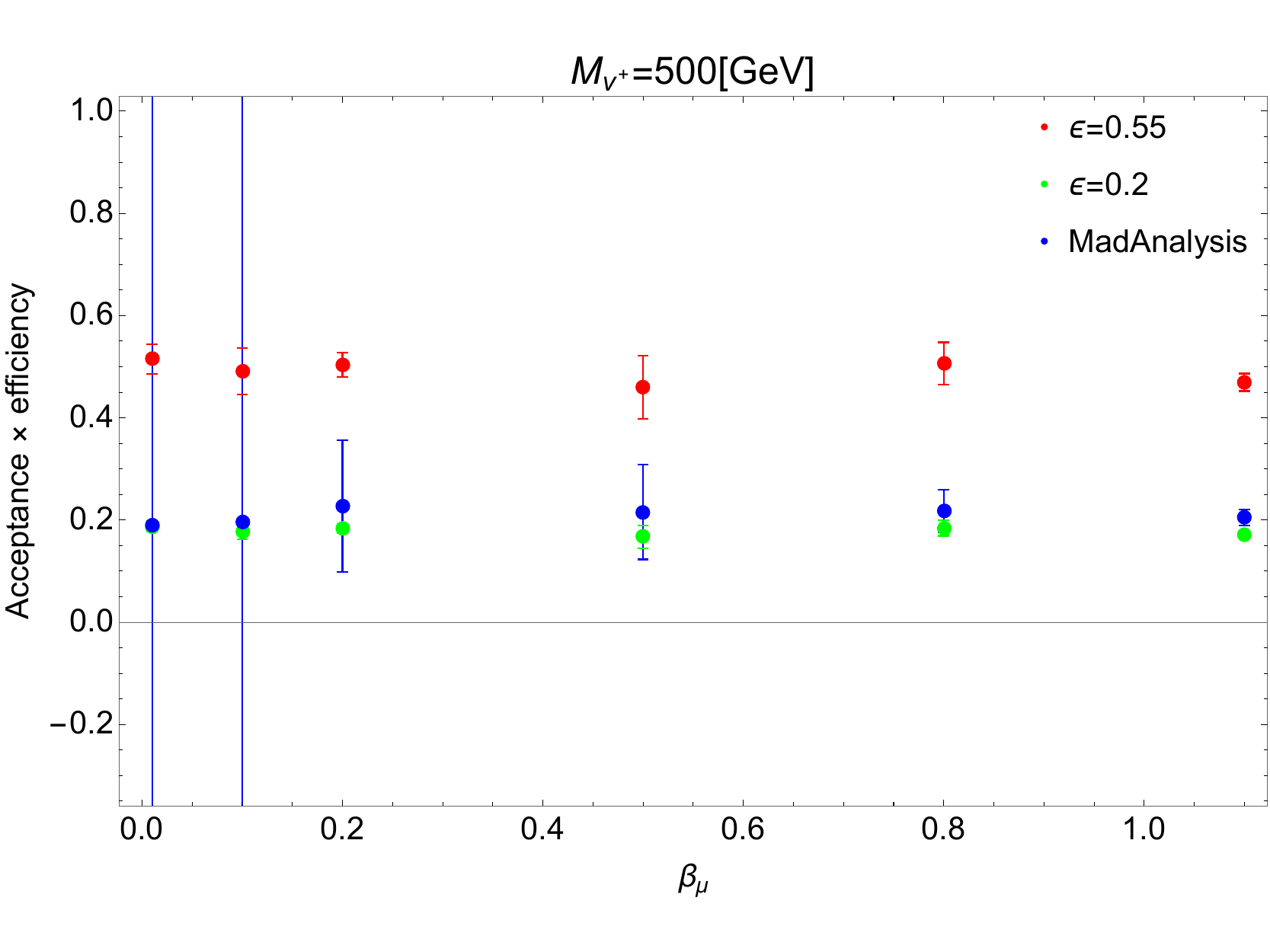}
    \caption{}
\end{subfigure}

    \begin{subfigure}{0.45\textwidth}
    \includegraphics[width=\textwidth]{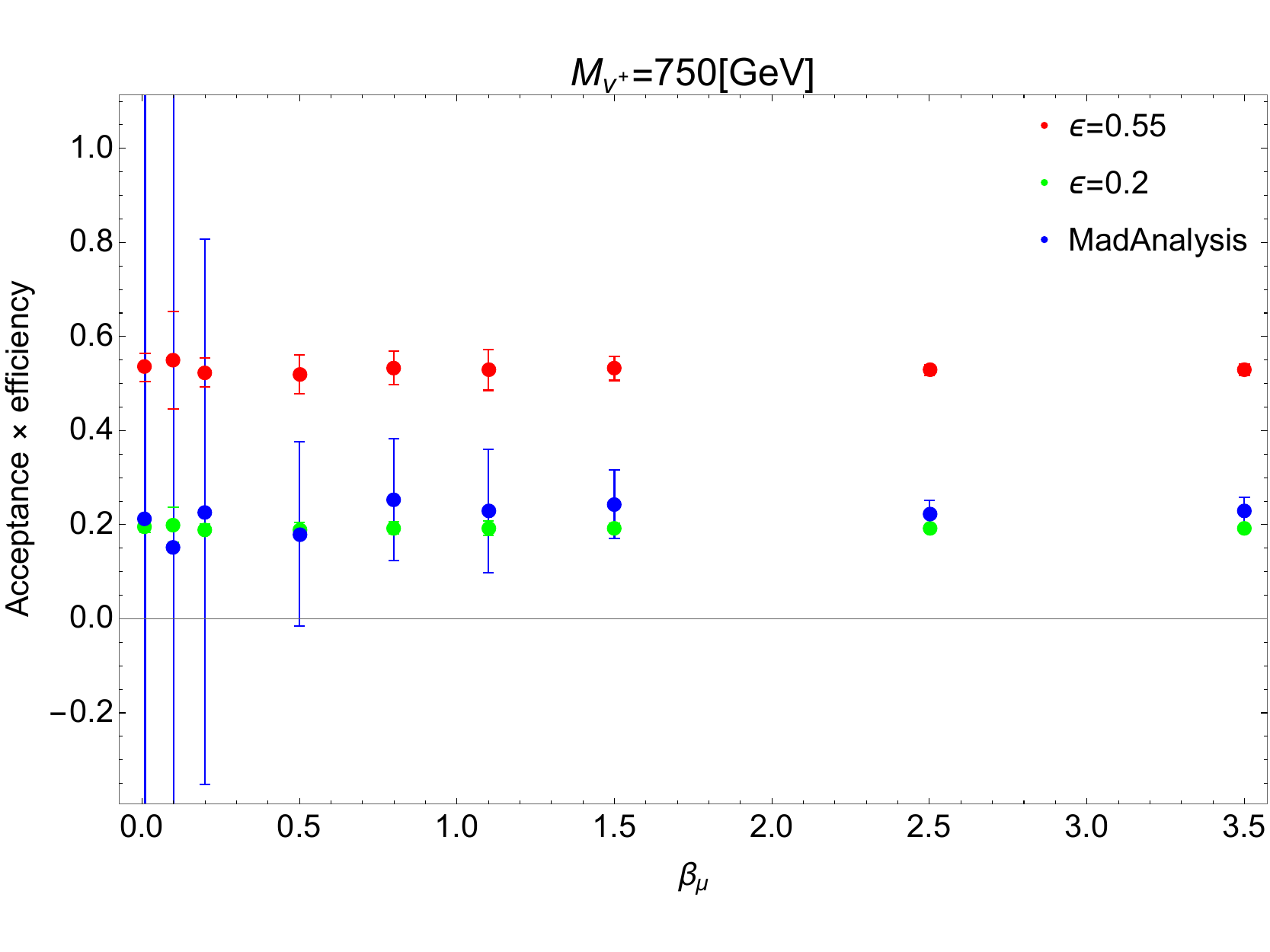}
    \caption{}
\end{subfigure}
\hspace{0.6cm}
\begin{subfigure}{0.45\textwidth}
    \centering
    \includegraphics[width=\textwidth]{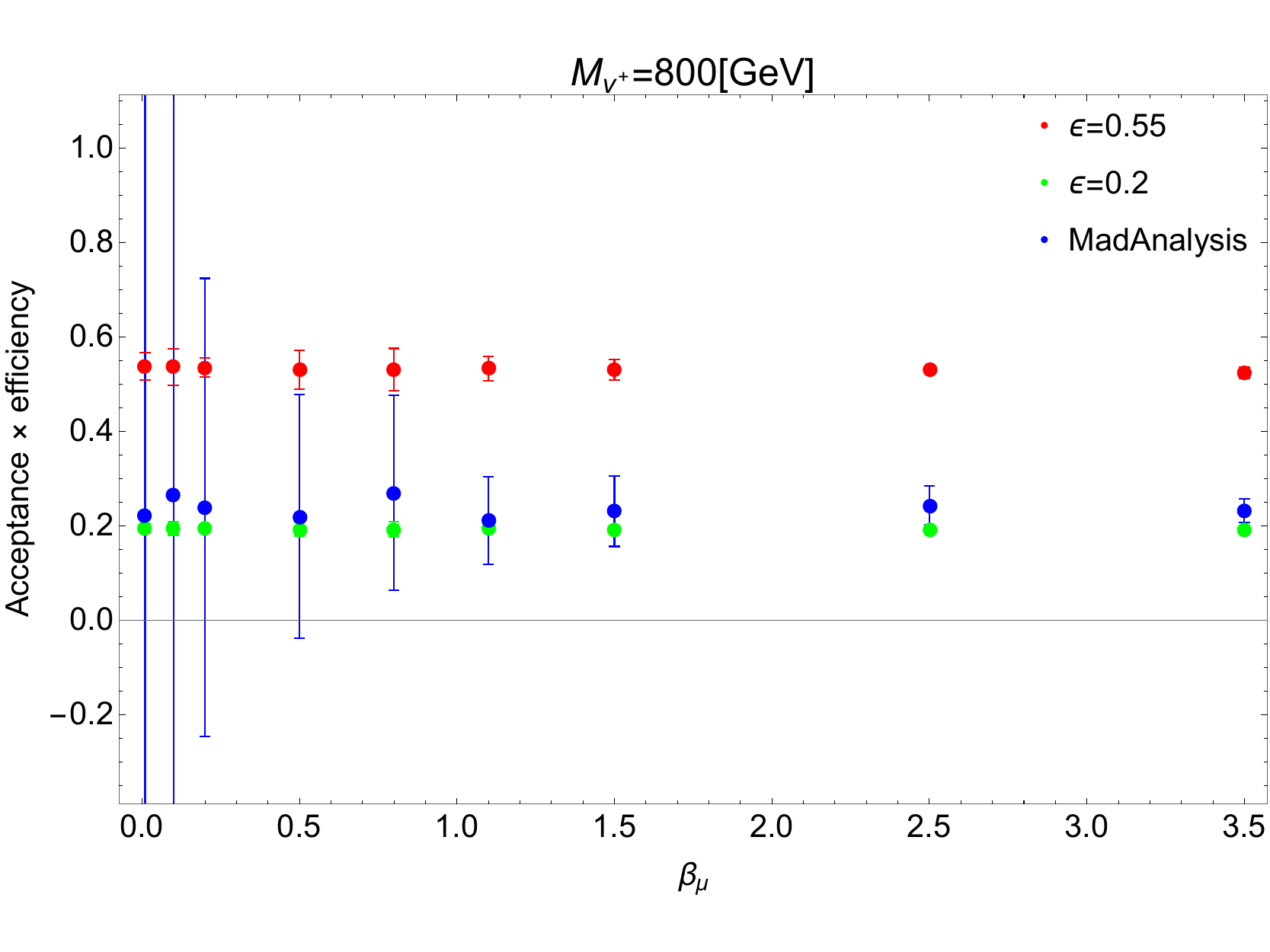}
    \caption{}
\end{subfigure}
\caption{Efficiency comparison between our cut-flow and the result obtained with MadAnalysis}
\label{val_masses}
\end{figure}

In order to check the consistency of our analysis, we validated our result with a full simulation using pythia 8.2, fastjet 3.3.4 and Delphes 3.5.0 \cite{pythia,fastjet,delphes}, the latter one interfaced with MadAnalysis 5  \cite{mad1,mad2,mad3,mad4,mad5,mad6,mad7,mad8,mad9} for recasting previous results from ATLAS. In particular, we considered the implementation of Ref. \cite{susylims_dy} that is publicly available in the MadAnalysis 5 Public Analysis Database \cite{DVN/EA4S4D_2020}. We focused on DY production and BP1. As can be seen from Figure \ref{val_masses}, the conservative choice of the efficiency shows a stronger concordance with the efficiencies obtained with MadAnalysis, however, considering a larger efficiency is helpful for future searches, considering the development of detector technology that will be implemented in future stages of the LHC. Therefore, we show our results considering both prescriptions for the detector efficiency.

\begin{figure}
    \centering
    \begin{subfigure}{0.45\textwidth}
    \includegraphics[width=\textwidth]{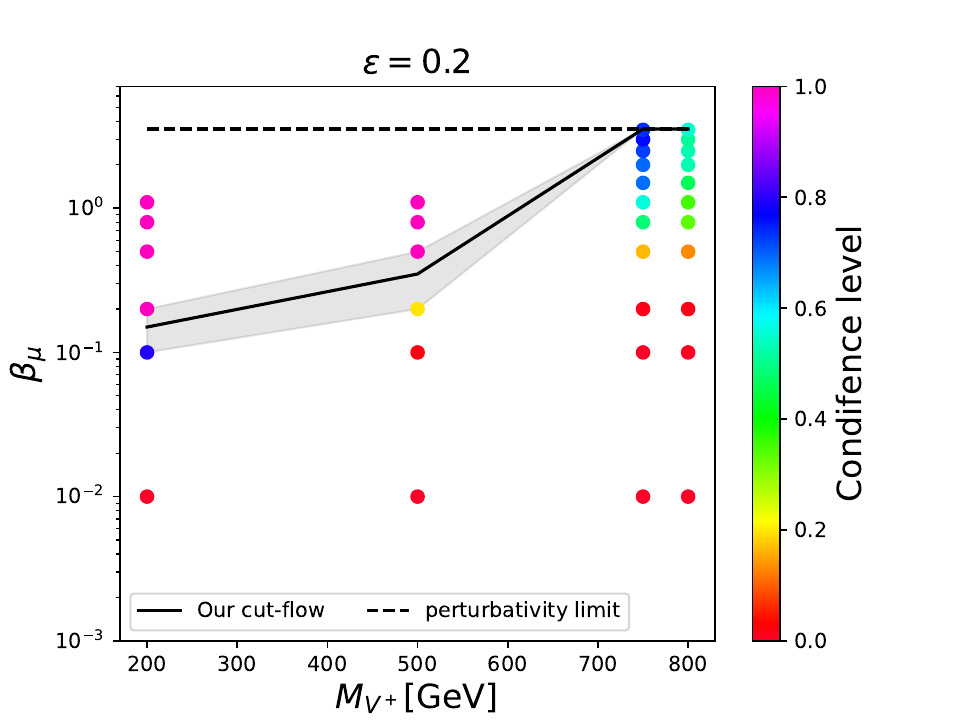}
    \caption{}
\end{subfigure}
\hspace{0.6cm}
\begin{subfigure}{0.45\textwidth}
    \centering
    \includegraphics[width=\textwidth]{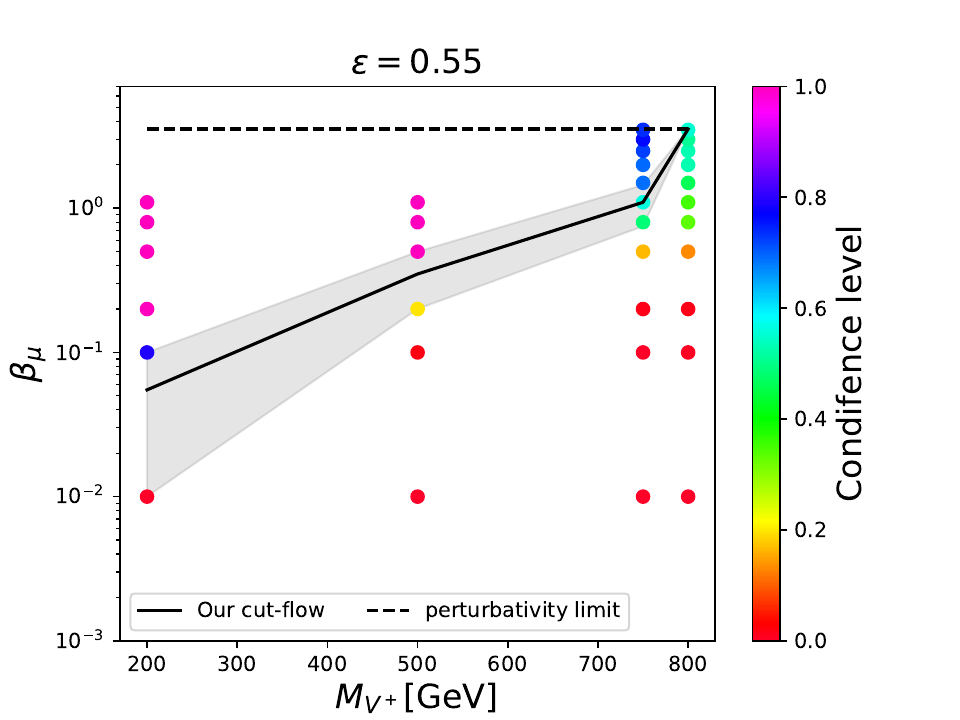}
    \caption{}
\end{subfigure}
    \caption{Comparison between the exclusion limits obtained from different methods.}
    \label{val_lims}
\end{figure}

\section*{Aknowledgements}
This work was funded by ANID - Millennium Program - ICN2019\_044. Also, we would like to thank to the DGIIP-UTFSM for funding during the development of this work. AZ was partially supported
by Proyecto ANID PIA/APOYO AFB220004 (Chile) and Fondecyt 1230110.

\bibliographystyle{utphys}
\bibliography{References}
\addcontentsline{toc}{chapter}{\bibname}

\end{document}